\documentclass[aps,prb,reprint,twocolumn,superscriptaddress]{revtex4-2}

\usepackage{graphicx}
\usepackage{bm}
\usepackage{hyperref}
\usepackage{amsmath}
\usepackage{amssymb}

\usepackage{comment}
\usepackage[table]{xcolor}
\usepackage{array}
\usepackage{multirow}
\usepackage{physics}
\usepackage{braket}

\usepackage{graphicx}
\usepackage{bm}
\usepackage{hyperref}
\usepackage{amsmath}
\usepackage{amssymb}
\usepackage{comment}

\usepackage[table]{xcolor}
\usepackage{array}
\usepackage{multirow}

\newcommand{\nset}{\{ n \}}
\newcommand{\rset}{\{\bm{r}\}}

\begin{document}

\title{Wavefunction approach to the fractional anomalous Hall crystal}
\author{Tixuan Tan}
\affiliation{Department of Physics, Stanford University, Stanford, CA 94305, USA}
\author{Julian May-Mann}
\affiliation{Department of Physics, Stanford University, Stanford, CA 94305, USA}
\author{Trithep Devakul}
\email{tdevakul@stanford.edu}
\affiliation{Department of Physics, Stanford University, Stanford, CA 94305, USA}

\begin{abstract}

We propose fractional anomalous Hall crystals (FAHCs) as possible ground states of strongly interacting electrons in parent bands with Berry curvature.
FAHCs are exotic states of matter that spontaneously break continuous translation symmetry to form a fractional Chern insulator.
We construct a unified family of variational wavefunctions that describe FAHCs and their competing states in the presence of uniform parent Berry curvature.
We calculate their variational energy with Coulomb interactions semi-analytically in the thermodynamic limit.
Our analysis reveals that FAHCs can be energetically favorable over both Wigner crystals and integer anomalous Hall crystals for sufficiently strong interactions or flat dispersion.

\end{abstract}

\maketitle

\textit{Introduction} --- 
Determining the fate of interacting electrons is a fundamental problem in the study of quantum matter.
Since the pioneering work of Wigner in the 1930s~\cite{wigner1934interaction}, it has been known that strong repulsive interactions can cause electrons to spontaneously break translation symmetry, forming a Wigner crystal (WC).
Nearly 50 years later, it was discovered that strongly interacting electrons in a magnetic field can form topologically ordered fractional quantum hall (FQH) states with robust ground-state degeneracies and fractionalized quasiparticles~\cite{tsui1982two}. 
In a magnetic field at low densities, FQH states often compete with WCs and the outcome of this competition depends on delicate energetic considerations~\cite{lam1984liquid, levesque1984crystallization, jiang1990quantum, he2005phase, archer2013competing}.

However, crystallization and topological order do not necessarily compete~\cite{PhysRevLett.56.873,PhysRevB.36.1620,PhysRevLett.57.922,PhysRevB.39.8525}. 
This is elegantly demonstrated by the concept of the ``Hall crystal'', where charge and topological order coexist in a magnetic field.
Although it is not the ground state with Coulomb interactions, the Hall crystal can emerge under suitably chosen interactions~\cite{PhysRevLett.57.922,PhysRevB.39.8525}.  


Recent advances have shown that Berry curvature and strong interactions can similarly lead to integer ``anomalous Hall crystals'' (IAHCs): states that spontaneously break continuous translation symmetry to form an insulator with Chern number $C\neq 0$ in the absence of a magnetic field. 
In the presence of a pinning potential, such a state would give rise to a quantized anomalous Hall effect.
The concept of IAHCs is motivated by recent experimental~\cite{aronson2024displacementfieldcontrolledfractionalchern,lu2023fractional,waters2024interplayelectroniccrystalsinteger,xie2024evenodddenominatorfractionalquantum,choi2024electricfieldcontrolsuperconductivity} and theoretical~\cite{
dong2023anomalous,zhou2024fractionalquantumanomaloushall,dong2023theoryfractionalquantumanomalous,guo2023theoryfractionalcherninsulator,yu2024moirefractionalcherninsulators,huang2024self,kwan2023moirefractionalcherninsulators,tan2024parent,PhysRevLett.132.236601,soejima2024anomaloushallcrystalsrhombohedral,dong2024stabilityanomaloushallcrystals,huang2024fractionalquantumanomaloushall,patri2024extendedquantumanomaloushall,sarma2024thermalcrossovercherninsulator}
 progress in rhombohedral multilayer graphene~\cite{chen2020tunable,Zhang2019nearly,Correlated2023Han,han2023large,Zhou_2021super,Zhou_2021half,de_la_Barrera_2022,Seiler_2022,Zhou_2022isospin,liu2023spontaneous,han2023orbital}.
Unlike the Hall crystal, where charge and topological order are merely compatible, the nonzero Chern number of the IAHC \emph{requires} translation symmetry breaking~\footnote{This conclusion follows from the fact that the IAHC can form at arbitrary densities of the parent band and the Lieb-Schultz-Mattis theorem~\cite{lieb1961two,oshikawa2000topological,hastings2004lieb}, if we assume the parent band has a fine underlying lattice}.
These topological states, enabled by spontaneous crystallization, are an exciting new frontier in the study of strongly correlated topological phases.  
\begin{figure}[t!]
    \centering
    \includegraphics[width=0.75\linewidth]{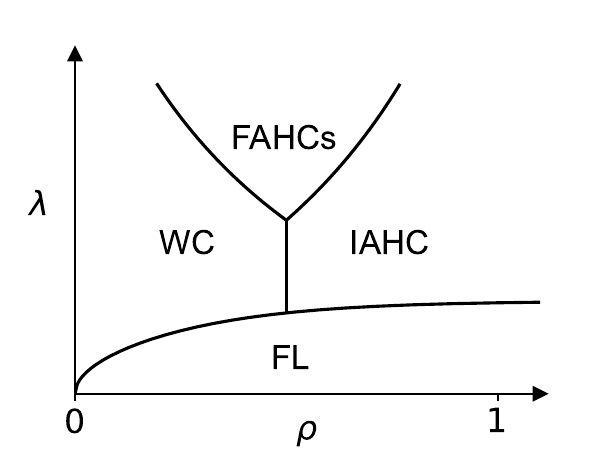}
    \caption{Proposed phase diagram as a function of the interaction strength $\lambda$ and dimensionless density $\rho$. 
    We identify a region where fractional anomalous Hall crystals (FAHCs) are energetically favorable ground states.
    Other phases shown are the Wigner crystal (WC), integer anomalous Hall crystal (IAHC), and Fermi liquid (FL). 
    }
    \label{fig_schematic}
\end{figure}

In this paper, we consider the possibility of a ``fractional anomalous Hall crystal'' (FAHC), an exotic state that spontaneously breaks continuous translation symmetry to form a fractional Chern insulator (FCI)~\cite{
Liu_2024,parameswaran2013fractional,bergholtz2013topological,PhysRevLett.106.236802,PhysRevLett.106.236804,PhysRevX.1.021014,sheng2011fractional}.
While there is no fundamental reason the FAHC cannot exist \emph{in principle}, it is unclear whether it can be stabilized as the ground state with realistic interactions, or if it requires specially engineered interactions similar to the Hall crystal~\cite{halperin1986compatibility,tevsanovic1989hall}.
Note that we focus on FAHCs that break \textit{continuous} translation symmetry (in contrast to similar states on a lattice~\cite{polshyn2022topological,sheng2024quantum,PhysRevB.109.115116,su2024generalized,gonçalves2024dopinginducedquantumanomaloushall,kourtis2014combined,kourtis2018symmetry}).
As we shall explain, this is an inherently difficult problem to solve.

The starting point is a continuum parent band with dispersion $E(\bm{k})$ (unbounded $\bm{k}$) and non-trivial Berry curvature and quantum geometry (encoded in the complex form factors $\mathcal{F}(\bm{k}^\prime,\bm{k})\equiv\bra{\bm{k}^\prime}e^{i(\bm{k}^\prime-\bm{k})\cdot\bm{r}}\ket{\bm{k}}$).
With sufficiently strong interactions and high parent Berry curvature, a possible ground state is the IAHC~\cite{dong2023anomalous,tan2024parent,soejima2024anomaloushallcrystalsrhombohedral,zhou2024fractionalquantumanomaloushall,PhysRevLett.132.236601,dong2024stabilityanomaloushallcrystals}.

The IAHC can be well understood within the framework of self-consistent Hartree-Fock (HF) theory.
In this framework, interactions generate a periodic HF potential, leading to the formation of HF bands, the lowest of which is fully filled.
When the filled band has Chern number $C\neq 0$, the resulting state is an IAHC.
By analogy to the FCI at fractional filling of a Chern band~\cite{
Liu_2024,parameswaran2013fractional,bergholtz2013topological,PhysRevLett.106.236802,PhysRevLett.106.236804,PhysRevX.1.021014,sheng2011fractional},
one might expect that a FAHC could arise at fractional filling of this HF band.
Herein lies the problem.
As the electron density decreases, rather than partially filling the HF band, it is more natural that the IAHC continuously adjusts its crystal unit cell size to maintain one electron per unit cell.
For the FAHC to form,  electrons would need to spontaneously crystallize into a lattice with the ``wrong'' period, corresponding to fractional $\nu=\frac{p}{q}$ electrons per unit cell.
Whether nature could favor such a state is a compelling prospect.

In general, controlled numerical simulation of this type of problem is highly challenging.
The unbounded momentum poses a significant issue for methods that rely on isolated bands within a Brillouin zone, such as exact diagonalization (ED) or density matrix renormalization group~\cite{soejima2020efficient}.
Additionally, the complex form factor introduces a sign problem for Monte Carlo techniques.
While HF theory can capture the WC or IAHC, it is fundamentally inadequate for describing highly entangled many-body states such as the FAHC.
Attempts at describing FAHC-like~\footnote{They are not true FAHCs since the alignment with the hBN provides a periodic moir\'e potential that fixes the unit cell.} states in rhombohedral graphene rely on a combination of HF and ED~\cite{dong2023anomalous,dong2023theoryfractionalquantumanomalous,guo2023theoryfractionalcherninsulator,zhou2024fractionalquantumanomaloushall,yu2024moirefractionalcherninsulators,huang2024self}, which is difficult to control~\cite{yu2024moirefractionalcherninsulators}.
The fact that the competing states have different (generally incommensurate) unit cells further complicates the task of analyzing their competition in the thermodynamic limit.

Given the combination of fundamental interest and inherent difficulty of this problem, we believe it is essential to focus on the simplest possible model, incorporating only the minimal essential ingredients.
To this end, we consider an ``ideal parent band'', with
uniform Berry curvature $\mathcal{B}$
and parabolic dispersion $E(\bm{k})=\frac{|\bm{k}|^2}{2 m_0}$. We add Coulomb interactions $V(\bm{r}) = \frac{V_{\mathrm{C}}}{2\pi|\bm{r}|}$ to generate correlated states. This leads to a characteristic length scale set by the Berry curvature $r_{\mathcal{B}}=\sqrt{\mathcal{B}}$, and an energy scale $E_{\mathcal{B}}=V(r_{\mathcal{B}})$. At electronic density $\rho_0$, the system is fully characterized by only two dimensionless parameters:
a coupling constant $\lambda\equiv m_0(E_{\mathcal{B}}r_{\mathcal{B}}^2)$ and a dimensionless density $\rho\equiv \rho_0 (2\pi r_{\mathcal{B}}^2)$.
The coupling constant $\lambda$ controls the relative strength of interactions to kinetic energy.

This model was recently proposed to study the IAHC~\cite{tan2024parent}.
At very low densities ($\rho \ll 1$), the Berry curvature becomes negligible, and the problem reduces to a standard 2D electron system with Wigner–Seitz radius $r_s\propto \lambda/\sqrt{\rho}$.  
For $\rho = 1$, the crystalline Brillouin zone (BZ) encloses exactly $2\pi$ parent Berry flux, and the $C=1$ IAHC is a favorable ground state for strong interactions~\cite{tan2024parent}.
Thus, the natural place to look for FAHCs is within the range $0<\rho<1$ and large $\lambda$.

We take a variational ansatz approach.
We construct a unified family of trial wavefunctions that describe the WC, IAHC, and FAHC states on equal footing.
Our FAHC wavefunctions are Laughlin-like~\cite{laughlin1983anomalous} FQH states, built via a mapping between the IAHC wavefunction and the lowest Landau level (LLL).
Unlike the standard Laughlin states, which have no tunable parameters,
our wavefunctions are parameterized by a single variable that optimizes the balance between kinetic and interaction energies.  
To a good approximation, the variational energies of these many-body wavefunctions can be calculated directly in the thermodynamic limit.

\begin{figure}[t!]
    \centering
    \includegraphics[width=1\linewidth]{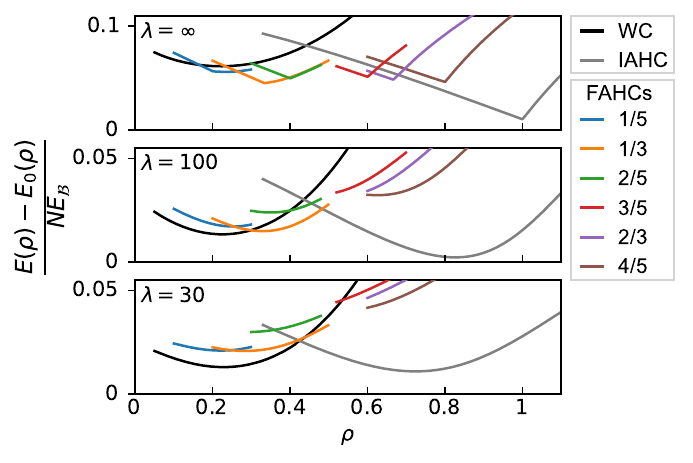}
    \caption{Energy per particle for the trial wavefunctions:  WC; $C=1$ IAHC; and FAHCs constructed from $\nu=p/q$ FQH states.
    The energy is shown as a function of $\rho$
    for $\lambda = 30$, $100$, and $10^7 \sim \infty$.
    A smooth offset $E_0(\rho)$
    has been subtracted for visibility (see Appendix for details). 
    }
    \label{fig_phase}
\end{figure}

Our findings are summarized by the phase diagram shown in Fig~\ref{fig_schematic}.
Our main result is the identification
of a region in phase space where the FAHC is energetically favorable over both the WC or IAHC.
This phase diagram highlights the importance of Berry curvature and strong interactions (or flat dispersion) in stabilizing FAHCs.
Fig~\ref{fig_phase} shows the energies of our WC, $C=1$ IAHC, and FAHC wavefunctions.

Some qualifications are in order.
Our results cannot rule out the existence of other competing states.  
Additionally, correlation effects beyond those captured by our trial wavefunctions may affect energetic competition.
However, in our approach, all competing states are treated on equal footing.
It is not clear \emph{a priori} whether such effects would favor or disfavor FAHCs relative to their competitors.
Further numerical and theoretical analysis is necessary to clarify this.
Nevertheless, our results establish that FAHCs can be a competitive ground state with Coulomb interactions.  

The remainder of this paper details the variational wavefunctions, and how we arrive at our result in Fig~\ref{fig_phase}.

\textit{Model} --- We consider the ideal parent band introduced in Ref.~\cite{tan2024parent}.
The key properties of this band are a parabolic dispersion $E(\bm{k})=\frac{|\bm{k}|^2}{2m_0}$ and the form factors
\begin{equation}
\mathcal{F}(\bm{k}^\prime,\bm{k})=\exp({-\tfrac{\mathcal{B}}{4}(|\bm{k}^\prime-\bm{k}|^2+2i\bm{k}^\prime\cross\bm{k})})
\label{eq:FormFact}
\end{equation}
where $\mathcal{B}$ is the parent Berry curvature.
The uniform quantum geometry can be understood by noting that these are the same form factors as the magnetic Bloch states of the LLL, except here we have a dispersive band with unbounded $\bm{k}$ (and hence an ``infinite'' total Chern number).
Since the second-quantized Hamiltonian, including density-density interactions, depends only on the form factors, the physics is fully determined at this point.

Nevertheless, to write down explicit first-quantized wavefunctions, we need a microscopic realization of this parent band.
We take the parent band wavefunctions~\cite{tan2024parent}
\begin{equation}
w_{\bm{k},n}(\bm{r})=
\langle\bm{r},n|\bm{k}\rangle
=\frac{e^{i\bm{k}\cdot\bm{r}}}{\sqrt{A}} \frac{e^{-\frac{\mathcal{B}}{4}|\bm{k}|^2}(\sqrt{\mathcal{B}}\bar{k})^n}{\sqrt{n!}}
\end{equation}
where $n=0,1,\dots,\infty$ is a microscopic orbital index, $\bar{k}=(k_x-ik_y)/\sqrt{2}$, and $A$ is the total system area (see Appendix for a derivation using a microscopic Hamiltonian). 
It is straightforward to verify that $\mathcal{F}(\bm{k}^\prime,\bm{k})=\sum_n \int d\bm{r} w^*_{\bm{k^\prime},n}(\bm{r})e^{i(\bm{k}^\prime-\bm{k})\cdot\bm{r}}w_{\bm{k},n}(\bm{r})$ gives rise to Eq~\ref{eq:FormFact}.
With this established, we can now discuss our trial wavefunctions.

\textit{Variational wavefunctions} --- 
To motivate our variational wavefunction, we first review the standard integer quantum Hall (IQH) and fractional quantum Hall (FQH) wavefunctions. 
The IQH wavefunction is a Slater determinant of LLL states.
In symmetric gauge, it can be written in terms of the Vandermonde determinant
\begin{equation}
    \Phi^{\mathrm{IQH}}(\rset) =  \prod_{i < j} (\bar{z}_i-\bar{z}_j) e^{-\frac{1}{2}\sum_i |z_i|^2/l^2_B}
\label{eq:IQHwf}
\end{equation}
where $\rset=\{\bm{r}_1,\dots,\bm{r}_N\}$ are the positions of all the $N$ particles, $z = \frac{x + i y}{\sqrt{2}}$, $\bar{z} = z^*$, and $l_B$ is the magnetic length. 
Equivalently, $\Phi^{\mathrm{IQH}}$ can be viewed as the Slater determinant of magnetic Bloch states $\phi^{\mathrm{LL}}_{\bm{k}}(\bm{r})$, with $\bm{k}$ in the magnetic BZ.

The paradigmatic FQH wavefunction is the Laughlin wavefunction~\cite{laughlin1983anomalous},
\begin{equation}\begin{split}
    \Phi^{\mathrm{FQH}}(\rset) &= (\bar{z}_i-\bar{z}_j)^{2s+1} e^{-\frac{1}{2} \sum_i |z_i|^2/l^2_B},
\label{eq:FQHwf}\end{split}\end{equation}
which describes a FQH state at filling $\nu=1/(2s+1)$.
Here and throughout, we leave the normalization of many-body wavefunctions implicit. 

We now turn our attention to AHCs in the ideal parent band. 
In the next section, we will introduce a trial wavefunction for the Chern number $C= 1$ IAHC state, valid at any density $\rho$.
Like the IQH, this IAHC state can be expressed as a Slater determinant of single-particle states $\psi^{\mathrm{IAHC}}_{\bm{k},n}(\bm{r})$, where $\bm{k}$ are crystal momenta and $n$ is the microscopic orbital index.
By utilizing the rich analytic structure of this wavefunction, we prove that
\begin{equation}
  \psi^{\mathrm{IAHC}}_{\bm{k},n}(\bm{r}) = \mathcal{N}_{\bm{k}}e^{\frac{\gamma}{4} \bm{\nabla}^2 } \Big[ \phi^{\mathrm{LL}}_{\bm{k}}(\bm{r}) \chi_n(\bm{r})  \Big]
\label{eq:ICBtoLLL}
\end{equation}
where $\phi^{\mathrm{LL}}_{\bm{k}}(\bm{r})$ are the LLL states.
Here, the magnetic BZ of the LLL is taken to match the BZ of the IAHC.
The function $\chi_n(\bm{r})$ is independent of $\bm{k}$ and encodes the orbital and crystal structure of the wave function.
The operator $e^{\frac{\gamma}{4}\bm{\nabla}^2}$ acting on $[\dots]$ has the effect of smearing the short-distance features, reducing contributions from rapidly varying components that cost high kinetic energy. 
The degree of smearing is controlled by the variational parameter $\gamma$. 

Using this result, we can now write the many-body wavefunction for the IAHC as
\begin{equation}
\Psi^{\mathrm{IAHC}}_{\nset}(\rset) \! = e^{\frac{\gamma}{4} \sum_{i} \bm{\nabla}^2_i} \left [ \Phi^{\mathrm{IQH}}(\rset) \prod_i \chi_{n_i}(\bm{r}_i)\right ],
\label{eq:IAHCwf}\end{equation}
where $\nset$ are the orbital indices of the different particles.
Given the mapping from the IQH to IAHC wavefunctions, the generalization to the FQH state is straightforward. For a given FQH wavefunction, we define the corresponding many-body FAHC wavefunction as
\begin{equation}
\Psi^{\mathrm{FAHC}}_{\nset}(\rset) = e^{\frac{\gamma}{4} \sum_{i} \bm{\nabla}^2_i} \left [\Phi^{\mathrm{FQH}}(\rset) \prod_i \chi_{n_i}(\bm{r}_i) \right ].
\label{eq:FAHCwf}\end{equation}
This construction can be used for any FQH state that lives in the LLL, such as the Laughlin state in Eq.~\ref{eq:FQHwf} or the Jain states~\cite{PhysRevLett.63.199}. 
This is our trial wavefunction. We shall return to the task of calculating the energy of this wavefunction later.

\textit{Crystal states and Landau level mapping} ---
When $\mathcal{B}=0$, our parent band reduces to a usual two-dimensional electron gas, where it is well established that significantly strong repulsive interactions lead to the formation of a WC. 
When $\mathcal{B}\neq 0$, 
one should also consider IAHCs with Chern number $C$. 
A wavefunction ansatz for the $C=1$ IAHC at $\rho=1$ was presented in Ref.~\cite{tan2024parent}.

We generalize this wavefunction to Chern $C$ crystals at arbitrary density $\rho$.
The wavefunction is a Slater determinant of the single-particle states
\begin{equation}
\begin{split}
\psi^{(C)}_{\bm{k},n}(\bm{r})&=\mathcal{N}_{\bm{k}}\sum_{\bm{g}} e^{-\frac{|\bm{k}+\bm{g}|^2}{4\xi^2}-i\pi C (\frac{\bm{k}\times \bm{g}}{\Omega}+\omega_{\bm{g}})}w_{{\bm{k}+\bm{g}},n}(\bm{r}).
\label{eq:IAHCansatz}\end{split}
\end{equation}
Here, $\bm{k}$ is the crystal momentum, $\bm{g}$ are reciprocal lattice vectors of the triangular crystal, written as $\bm{g} = n_1 \bm{b}_1 + n_2 \bm{b}_2$ with $n_1,n_2\in \mathbb{Z}$, and $\omega_{\bm{g}}=(n_1-1)(n_2-1)$. $\Omega = |\bm{b}_1 \times \bm{b}_2|$ is the area of the crystal BZ. 
This crystal has one electron per unit cell, so the density is $\rho_0 = \Omega/4\pi^2$ (equiv. $\rho=\mathcal{B}\Omega/2\pi$). 
In our forthcoming discussions, it is useful to think of $\Omega$ as parameterizing the size of the crystal unit cell. 
The parameter $\xi$ is the single variational parameter of the wavefunction (from which $\gamma$ in  Eq.~\ref{eq:ICBtoLLL} will be derived). 
The momentum dependent normalization factor is $\mathcal{N}_{\bm{k}} = \mathcal{N} + \mathcal{O}(\epsilon)$, where
$\mathcal{N}=\sqrt{\Omega/(2\pi\xi^2 A)}$ and $\epsilon\equiv e^{-4\pi^2 \xi^2/\sqrt{3\Omega^2}}$.
Explicitly, all momentum-dependent terms are $\mathcal{O}(\epsilon)$. 
When $\epsilon$ is small (as in the cases considered), we can simplify Eq.~\ref{eq:IAHCansatz} by taking $\mathcal{N}_{\bm{k}} \approx \mathcal{N}$. 
This approximation greatly simplifies the calculation of energy in the thermodynamic limit in the next section.


When $C = 0$, Eq.~\ref{eq:IAHCansatz} describes a WC state, while the $C\neq 0$ cases are the Chern number $C$ IAHCs. 
The charge density of the WC is strongly modulated. 
In contrast, the IAHC has almost uniform charge density (fluctuations are $\mathcal{O}(\epsilon$)), but periodically modulated orbital texture.
In this work, we are primarily interested in the $C = 1$ IAHC, which we will refer to as ``the IAHC'' from here on.  
In the Appendix, we further justify the wavefunction ansatz Eq~\ref{eq:IAHCansatz} by showing that it is a local minimum of the energy functional, within the space of Slater determinant states, at large $\lambda$ near $\rho \approx 0$ and $\rho \approx 1$ for the WC and IAHC respectively.

We now establish the connection between the IAHC and the LLL. 
This finding connects to a broader body of work on so-called ``ideal Chern bands''~\cite{parameswaran2013fractional,Roy2014bandgeometry,wang2021exact,ledwith2022vortexability,claassen2015position,Ledwith2020fractionalchern,Lee2017engineering,jackson2015geometric,BrunoKahler2021}.
To do this, we rewrite the single-particle state $\psi^{\mathrm{IAHC}}_{\bm{k},n}(\bm{r})\equiv \psi^{(1)}_{\bm{k},n}(\bm{r})$ in Eq.~\ref{eq:IAHCansatz} in terms of a broadening operator acting on a generating function $\varphi_{\bm{k},n}(\bm{r})$,
\begin{equation}\begin{split}
&\psi_{\bm{k},n}^{\mathrm{IAHC}}(\bm{r}) =  \mathcal{N}_{\bm{k}} e^{\frac{\gamma}{4}\bm{\nabla}^2}\varphi_{\bm{k},n}(\bm{r}),\phantom{=} \gamma \equiv \tfrac{1}{\xi^2} + \mathcal{B}-\tfrac{2\pi}{\Omega}.
\end{split}\label{eq:BroadGenerating}\end{equation}
We have chosen this specific definition of $\gamma$ because it allows us to write $\varphi_{\bm{k},n}$ as
\begin{equation}
    \begin{split}
         &\varphi_{\bm{k},n}(\bm{r}) = e^{i\bm{k}\cdot\bm{r}} e^{-\frac{\pi}{2\Omega}|\bm{k}|^2}  \varphi^A_{\bar{k},n}(\bm{r})
    \end{split}
\end{equation}
where $\varphi^A_{\bar{k},n}(\bm{r})$ is an antiholomorphic function of momentum $k$. 
It can be verified that $\varphi^A_{\bar{k},n}(\bm{r})$  satisfies the momentum space boundary conditions 
\begin{equation}
    \varphi^A_{\bar{k}+\bar{g},n}(\bm{r}) = e^{-i \bm{g}\cdot \bm{r}  + i\pi (1-\omega_{\bm{g}}) + \frac{\pi}{\Omega} (\bar{g}+2\bar{k}) g } \varphi^A_{\bar{k},n}(\bm{r}).
\label{eq:magneticBC}\end{equation}

Since $\varphi^A_{\bar{k},n}$ is antiholomorphic, its momentum dependence is uniquely determined by its boundary conditions. Specifically, the boundary conditions in Eq.~\ref{eq:magneticBC} indicate that $\varphi^A_{\bar{k},n}$ has the same momentum dependence as the states in LLL~\cite{wang2021exact, wang2023origin,PhysRevB.108.205144}. 
We can therefore write (see Appendix) 
\begin{equation}
    \varphi_{\bm{k},n}(\bm{r}) = \phi^{\mathrm{LL}}_{\bm{k}}(\bm{r})\chi_n(\bm{r})
\label{eq:GeneratorToLL}
\end{equation}
where $\chi_n(\bm{r})$ is a momentum independent function and $\phi^{\mathrm{LL}}_{\bm{k}}(\bm{r})$ is the magnetic Bloch states in the LLL with magnetic length $l_B = \sqrt{2\pi/\Omega}$.
We have therefore arrived at Eq~\ref{eq:ICBtoLLL}.
Taking the Slater determinant, we arrive at the many-body wavefunction $\Psi^{\mathrm{IAHC}}$ in Eq.~\ref{eq:IAHCwf}.

The generalization to the FAHC wavefunction in Eq.~\ref{eq:FAHCwf} is straightforward, as we can simply replace the IQH wavefunction with a FQH wavefunction.
For a FQH state at filling $0<\nu<1$ of the LLL, the FAHC wavefunction occurs at density $\rho_0^{\text{FAHC}} = \nu\Omega/4\pi^2 = \nu \rho_0^{\text{IAHC}}$, where $\rho_0^{\text{IAHC}}$ is the density of the IAHC with the same crystal lattice (i.e. same value of $\Omega$). In this way, we can heuristically understand the FAHC as arising from fractional filling of the IAHC.


\textit{Variational energies} --- 
To determine the energetic competition between the WC, IAHC, and FAHCs, we calculate their variational energies in the presence of Coulomb interactions $V(\bm{r})=\frac{V_{\mathrm{C}}}{2\pi|\bm{r}|}$.
The total interaction energy can be written in terms of the density-density correlation function
\begin{equation}
E^{\mathrm{int}} = \frac{1}{2}\int d\bm{r}d\bm{r}^\prime V(\bm{r}-\bm{r}^\prime)\langle {:}\hat{\rho}(\bm{r}) \hat{\rho}(\bm{r}^\prime){:}\rangle
\end{equation}
where $\hat{\rho}(\bm{r})$ is the charge density operator and ${:}\hat{\mathcal{O}} {:}$ denotes normal ordering.

To calculate the interaction energy of the WC, we use the fact that Eq.~\ref{eq:IAHCansatz} with $C=0$ describes an array of Gaussian-localized wavepackets with localization length $\ell_{\mathrm{WC}} = (1+\mathcal{B}\xi^2)/\xi$. When $\ell_{\mathrm{WC}}$ is small compared to the WC lattice constant $a_0 = \sqrt{2/(\sqrt{3}\rho_0)}$, the energy per particle is 
\begin{equation}
\frac{E^{\mathrm{int}}_{\mathrm{WC}}}{N}=\frac{E_\mathcal{B}r_\mathcal{B}}{2a_0}\left[A(\tfrac{1}{2})+\frac{A(\frac{3}{2})\ell_{\mathrm{WC}}^2}{4a_0^2} + \cdots\right]
\end{equation}
where $A(s)\equiv 6\zeta(s)(\sum_{n=0}^{\infty} (3n+1)^{-s}-(3n+2)^{-s})$, $\zeta(s)$ is the Riemann zeta function,
 and $\cdots$ indicates terms that are further suppressed by powers $\ell_{\mathrm{WC}}/a_0$ (see Appendix for further details).

Next, we consider the interaction energies of the IAHC and FAHCs.
To calculate these, we consider the pair correlation function 
$g(\bm{r},\bm{r}^\prime)\equiv \frac{1}{\rho_0^2}\langle{:}\hat{\rho}(\bm{r})\hat{\rho}(\bm{r}^\prime){:}\rangle$.
Defining the Fourier transform $g_1(\bm{q})=\frac{1}{A}\int d\bm{r}d\bm{r}^\prime e^{-i\bm{q}\cdot(\bm{r}-\bm{r}^\prime)}[g(\bm{r},\bm{r}^\prime)-1]$ 
\cite{PhysRevB.30.558,Jain_2007}, we have
\begin{equation}
\begin{split}
\frac{E^{\mathrm{int}}_{\mathrm{AHC}}}{N}
= \frac{\rho_0}{2} \int \frac{d\bm{q}}{4\pi^2} g_1^{\mathrm{AHC}}(\bm{q})V(\bm{q}),
\end{split}
\end{equation}
where $V(\bm{q})=V_C/|\bm{q}|$. 
We now relate the pair correlation functions of the IAHC and FAHC to those of the corresponding QH states. 
Up to terms $\mathcal{O}(\epsilon)$, we have (see Appendix for details)
\begin{equation}
g_1^{\mathrm{AHC}}(\bm{q})=e^{-\frac{1}{4}\gamma^2 \xi^2|\bm{q}|^2 } g^{\mathrm{QH}}_1(\bm{q})
\label{eq:g1map}\end{equation}
where $g^{\mathrm{QH}}_1(\bm{q})$ is calculated for the QH state with magnetic length 
$l_B = \sqrt{2\pi/\Omega}$.
Therefore, these interaction energies can be thought of as the energy of the QH state with an effective interaction $V_{\mathrm{eff}}(\bm{q})=V(\bm{q})e^{-\frac{1}{4}\gamma^2\xi^2|\bm{q}|^2}$ 
(note that $\gamma$ is a function of $\xi$, Eq~\ref{eq:BroadGenerating}).

The kinetic energy of all the states (WC and AHCs) is simply
\begin{equation}
\frac{E^{\mathrm{kin}}}{N}=
\frac{1}{N}\sum_i\langle\frac{-\bm{\nabla}_i^2}{2m_0}\rangle
=\frac{\xi^2}{m_0}.
\end{equation}
up to $\mathcal{O}(\epsilon)$.
The total energy, $E=E^{\mathrm{kin}}+E^{\mathrm{int}}$, should then be minimized with respect to the variational parameter $\xi$.

The energy of the IAHC is of a simple analytic form. 
The IQH pair correlation function is $g^{\mathrm{IQH}}_1(\bm{q})=-2\pi l_B^2 e^{-\frac{1}{2}l_B^2|\bm{q}|^2}$\cite{jancovici1981exact, PhysRevB.33.2481}, which gives
\begin{equation}
\frac{E_{\mathrm{IAHC}}}{N}
=\frac{\xi^2}{m_0} -\frac{V_{\mathrm{C}}}{4\sqrt{\pi}}\frac{1}{\sqrt{\gamma^2\xi^2+4\pi/\Omega}}
\end{equation}
At $\rho=1$, this reduces to the expression derived in Ref.~\cite{tan2024parent} using a different method.

For the FAHC, 
we utilize a representation for $g_1^{\mathrm{FQH}}(\bm{q})$ of the form proposed by Girvin~\cite{PhysRevB.30.558},
\begin{equation}
\begin{split}
g_1^{\text{FQH}}(\bm{q}) =& -2\pi l_B^2e^{-\frac{1}{2}l_B^2 |\bm{q}|^2}\\
&+\sum_{m\in \mathrm{odd}}8\pi l_B^2c_m L_m(l_B^2 |\bm{q}|^2)e^{-l_B^2 |\bm{q}|^2}
\end{split}
\end{equation}
where $L_m(x)$ are Laguerre polynomials and $c_m$ are fitted coefficients that depend on the FQH state, listed in the Appendix.

The toal energies (minimized with respect to $\xi$) of the different crystals are shown in Fig.~\ref{fig_phase} as a function of $\rho$ for a selection of $\lambda$.
The error in calculating these energies is controlled by the parameter $\epsilon$, which we find is small compared to their energy differences.
Further details of the calculation and error analysis are given in the Appendix.

\textit{Discussion and conclusions} --- 
It is interesting to note from Fig~\ref{fig_phase} that the energy of the IAHC exhibits a cusp at $\rho=1$ for $\lambda\rightarrow\infty$.  
Indeed, an analytic expression can be obtained
\begin{equation}
\lim_{\lambda\rightarrow\infty} \frac{E_{\mathrm{IAHC}}}{N} = -\frac{E_{\mathcal{B}}}{2\sqrt{2}}{\frac{\sqrt{\pi}}{\sqrt{1+|1-\rho^{-1}|}}}.
\end{equation}
This conforms to the intuition that the IAHC is most stable when the parent Berry flux enclosed in the crystal BZ is $2\pi$.
That is, there is an energy cost associated with ``rounding'' the enclosed Berry flux to $2\pi$~\cite{dong2024stabilityanomaloushallcrystals}.
Interestingly, this cusp suggests that the IAHC should appear incompressible in this limit, as the chemical potential $\mu\propto dE/d\rho$ exhibits a discontinuity.
This cusp is smoothed for any non-infinite $\lambda$, and we believe that non-uniform Berry curvature or quantum geometry will also have similar effects.



Determining the true phase diagram of interacting electrons in the presence of Berry curvature remains an important numerical challenge.
Our results suggest the phase diagram sketched in Fig~\ref{fig_schematic}, where FAHCs are energetically favorable in a density window between the WC and IAHC for sufficiently strong interactions.
However, the precise details, such as phase boundaries and exact locations of specific FAHCs, should not be interpreted literally from our findings.
Correlations effects that go beyond our wavefunctions, such as fluctuations of the low-energy Goldstone modes associated with the crystal lattice, could affect the competition in unknown ways.
Moreover, other competing states (e.g. stripes, bubbles, or other exotic phases) could emerge that we did not consider.
Our findings underscore the need for new theoretical approaches to address these questions in a controlled manner.

More significantly, we have demonstrated that FAHCs are a competitive ground state with realistic interactions.
This raises intriguing questions about the fundamental physical properties of FAHCs.
Much remains to be explored regarding the general transport properties and the effect of disorder on topological crystalline states~\cite{patri2024extendedquantumanomaloushall,sarma2024thermalcrossovercherninsulator}.
Additionally, the collective modes of the FAHCs are of particular interest. 
Two gapless Goldstone modes are guaranteed~\cite{watanabe2020counting},  but it would be interesting to determine if the neutral mode spectrum has a roton-like minimum~\cite{PhysRevB.33.2481} similar to that observed in FQH states.
Our results lay the groundwork for future investigations into the fascinating properties of this novel phase of matter.


\textit{Acknowledgements} ---
We thank Junkai Dong, Tomohiro Soejima, Aidan P. Reddy, Dan Parker, and Sid Parameswaran for helpful conversations. TT is supported by the Stanford Graduate Fellowship.
TD acknowledges support from a startup fund
at Stanford University.

\bibliography{FAHC.bib}

\appendix 
\onecolumngrid
\section{Infinite Chern band with Constant Berry Curvature}\label{sup:ICB}
To realize an ideal parent band, we use the microscopic``infinite Chern band'' (ICB) model introduced in Ref.~\cite{tan2024parent}. The starting point is a system of electrons with infinitely many orbital degrees of freedom. The Hamiltonian for this system is, in second quantized form,

\begin{equation}\begin{split}
&\hat{\mathcal{H}}^{\mathrm{ICB}}(\bm{k}) =  c^\dagger_{\bm{k},n'} H^{\mathrm{ICB}}_{n^\prime n}(\bm{k})c_{\bm{k},n}\\
&H^{\mathrm{ICB}}_{n^\prime n}(\bm{k}) = \frac{|\bm{k}|^2}{2m_0}\delta_{n^\prime n} + J L_{n^\prime n}(\bm{k})
\end{split}\label{eq:ICB}\end{equation}
where $n \geq 0$ indexes the microscopic orbitals, $c^\dagger_{\bm{k},n}$ is the electronic creation operator for orbital $n$, and
\begin{equation}
  L_{n^\prime n}(\bm{k})= \sum_{n^{\prime\prime}}(b^\dagger_{n^\prime n^{\prime\prime}} -\sqrt{\mathcal{B}}k \delta_{n^\prime n^{\prime\prime}})(b_{n^{\prime\prime}  n}-\sqrt{\mathcal{B}}\bar{k} \delta_{ n^{\prime\prime} n }).
\end{equation} 
where $k \equiv (k_x+ik_y)/\sqrt{2}$ and $\bar{k} \equiv k^*$. Here and throughout, we will use bold symbols for momentum vectors, and unbolded symbols for complex momentum. The matrix $b_{{n^\prime n}} = \sqrt{n+1}\delta_{n^\prime,n+1}$ and $b^\dagger_{{n^\prime n}} = \sqrt{n}\delta_{n^\prime,n-1}$ couples different orbitals, and can be understood as the matrix representation of the simple harmonic oscillator creation and annihilation operators: $b_{{n^\prime n}} = \bra{n'} b \ket{n}$, where $\ket{n}$ and $b$ are SHO states and ladder operators respectively. The parameter $\mathcal{B}$ controls the Berry curvature of the bands. 

The ICB is the lowest energy band of Eq.~\ref{eq:ICB} when $J > 0$. The wavefunction of the ICB state with momentum $\bm{k}$ has the form, 
\begin{equation}
w_{\bm{k},n}(\bm{r})=\braket{\bm{r},n|\bm{k}}=\exp \left(i\bm{k}\cdot\bm{r} \right)\braket{n|s^{\mathcal{B}}_{\bm{k}}} = \exp \left (i\bm{k}\cdot\bm{r} - \frac{\mathcal{B}}{4} |\bm{k}|^2\right)\frac{(\sqrt{\mathcal{B}} \text{ } \bar{k})^n}{\sqrt{n!}}
\label{eq:ICBStates}\end{equation}
where we have defined the spinor,
\begin{equation}\begin{split}
\ket{s^{\mathcal{B}}_{\bm{k}}} &=  \exp \left(\sqrt{\mathcal{B}}(\bar{k}b^\dagger-kb)\right) \ket{0}
\end{split}\end{equation}
The ICB has a simple quadratic dispersion,  $E(\bm{k}) = \frac{|\bm{k}|^2}{2m_0}$. Higher energy bands are separated from the ICB by integer multiples of $J$. We consider the limit $J\rightarrow \infty$, such that the ICB is well separated from the higher energy bands of Eq.~\ref{eq:ICB}. 

The geometric properties of the ICB can be understood from its form factors,
\begin{equation}\begin{split}
\mathcal{F}_{\mathcal{B}}(\bm{k}^\prime,\bm{k}) &= \bra{\bm{k}^\prime}\exp \left(i(\bm{k}^\prime-\bm{k})\cdot\bm{r}\right)\ket{\bm{k}} \\ &=\exp\left(-\frac{\mathcal{B}}{4}(|\bm{k}^\prime-\bm{k}|^2+2i\bm{k}^\prime\times\bm{k})\right).
\label{eq:FormFactSM}\end{split}\end{equation}
These are the same form factors as the magnetic Bloch states for lowest Landau level (LLL) on a torus, except here we have an unbounded $\bm{k}$. Taking derivatives of Eq.~\ref{eq:FormFactSM}, we verify that the Berry curvature is constant and equal to $\mathcal{B}$. 

\section{Anti-Holomorphic wavefunction}

The $C = 1$ integer anomalous Hall crystal wavefunction is the Slater determinant of the following single-particle Bloch states $\ket{\bm{k}^{(1)}}$

\begin{equation}
\begin{split}
\braket{\bm{r},n|\bm{k}^{(1)}}=\psi^{(1)}_{\bm{k},n}(\bm{r})&=\mathcal{N}_{\bm{k}}\sum_{\bm{g}} \exp\left(-\frac{|\bm{k}+\bm{g}|^2}{4\xi^2}-i\pi  (\frac{\bm{k}\times \bm{g}}{\Omega}+\omega_{\bm{g}})\right) w_{\bm{k}+\bm{g},n}(\bm{r})
\end{split}
\end{equation}
where $\bm{k}$ is the crystal momentum of the single particle state, $n$ is the orbital index, $\Omega$ is the Brillouin zone (BZ) area, $\xi$ is the variational parameter, and $\mathcal{N}_{\bm{k}}$ is the normalization factor. The sum is over reciprocal lattice vectors $\bm{g}=n_1\bm{g}_1+n_2\bm{g}_2$ for $n_1,n_2\in \mathbb{Z}$, where $\bm{g}_1$ and $\bm{g}_2$ are the primitive reciprocal lattice vectors of the triangular lattice. The phase factor $\omega_{\bm{g}}$ is defined as $\omega_{\bm{g}}\equiv(n_1-1)(n_2-1)$.  We use the superscript $(1)$ to distinguish this wavefunction from the wavefunctions we will use later for the Wigner Crystal.

To proceed, it will be useful to define new parameters, $\delta$ and $\gamma$,
\begin{equation}
\delta \equiv \frac{2\pi}{\Omega}-\mathcal{B}, \phantom{=} \gamma\equiv \frac{1}{\xi^2}+\delta,
\end{equation}
and write $\psi^{(1)}_{\bm{k},n}(\bm{r})$ in terms of a broadening operator $\exp \left (\frac{\gamma\bm{\nabla}^2}{4}\right )$ acting on a generating function $\varphi_{\bm{k},n}(\bm{r})$,
\begin{equation}
\begin{split}
\psi^{(1)}_{\bm{k},n}(\bm{r})&=\mathcal{N}_{\bm{k}} \exp \left(\frac{\gamma\bm{\nabla}^2}{4}\right)\left [\sum_{\bm{g}} \exp\left(\frac{\delta |\bm{k}+\bm{g}|^2}{4}-i\pi  (\frac{\bm{k}\times \bm{g}}{\Omega}+\omega_{\bm{g}})\right) w_{\bm{k}+\bm{g},n}(\bm{r}) 
\right ]\\
&\equiv \mathcal{N}_{\bm{k}} \exp \left(\frac{\gamma\bm{\nabla}^2}{4}\right)\varphi_{\bm{k},n}(\bm{r})
\end{split}
\end{equation}
The generating function $\varphi_{\bm{k},n}(\bm{r})$ is anti-holomorphic in $k$ up to an overall factor of $ \exp \left(-\frac{\pi k\bar{k}}{\Omega} + i \bm{k}\cdot \bm{r}\right) $,

\begin{equation}\label{eq:explicitphi}
\begin{split}
\varphi_{\bm{k},n}(\bm{r})&=\exp \left(-\frac{\pi k\bar{k}+i\bm{k}\cdot\bm{r}}{\Omega}\right)\sum_{\bm{g}}\exp \left(-i\pi \omega_{\bm{g}}\right)\exp \left(i(g\bar{z}+z\bar{g})\right)\exp \left(-\frac{\pi(g\bar{g}+2\bar{k}g)}{\Omega}\right) \frac{(\sqrt{\mathcal{B}}(\bar{k}+\bar{g}))^n}{\sqrt{n!}}\\
&\equiv \exp \left(-\frac{\pi k\bar{k}}{\Omega}+i\bm{k}\cdot\bm{r}\right) \varphi^A_{\bar{k},n}(\bm{r})\\
\end{split}
\end{equation}
where $ z\equiv \frac{x+iy}{\sqrt{2}}$. We use the notation $ \varphi^A_{\bar{k},n}(\bm{r})$ to stress the fact that it is anti-holomorphic in $k$.  Furthermore,  $ \varphi^A_{\bar{k},n}(\bm{r})$ satisfies the following $\bm{k}$-space boundary condition.

\begin{equation}
\begin{split}
\varphi^A_{\bar{k}+\bar{G},n}(\bm{r})
=-\exp \left(-i\bm{G}\cdot\bm{r}\right) \varphi^A_{\bar{k},n}(\bm{r}) \exp \left(\frac{\pi}{\Omega}(2\bar{k}G+\bar{G}G)\right)
\label{eq:GenFuncBC}\end{split}
\end{equation}
where $\bm{G}$ is a primitive reciprocal lattice vector. Anti-holomorphic functions are uniquely determined (up to an overall constant) by their boundary condition. The modified Weierstrass sigma function with periods $\bar{a}_1$ and $\bar{a}_2$\footnote{The corresponding Mathematica representation is $\sigma(\bar{z})=\text{WeierstrassSigma}\left[\bar{z},\text{WeierstrassInvariants}\left[\left\{\frac{\bar{a}_1}{2},\frac{\bar{a}_2}{2}\right\}\right]\right]$}, abbreviated as $\sigma(\bar{z})$ below, is anti-holomorphic in $\bar{z}$ and satisfies boundary conditions\cite{wang2021exact,Haldane_2018}
\begin{equation}
\sigma(\bar{z}+\bar{a}_i)=-\exp \left(a_i(\bar{z}+\frac{\bar{a}_i}{2})\right)\sigma(\bar{z})
\label{eq:WsfBC}\end{equation}
where $\bm{a}_i$ are the primitive lattice vectors of a given lattice. Based on Eq.~\ref{eq:WsfBC}, $\varphi^A_{\bar{k},n}(\bm{r})$ can be written as
\begin{equation}
\varphi^A_{\bar{k},n}(\bm{r})=\sigma\left(\bar{z}/\sqrt{\frac{2\pi}{\Omega}}+i\bar{k}\sqrt{\frac{2\pi}{\Omega}}\right)\exp \left(-iz\bar{k}\right)\exp \left(-\frac{z\bar{z}\Omega}{4\pi}\right)\chi_n(\bm{r}),
\end{equation}
where $\chi_n(\bm{r})$ is $\bm{k}$-independent function of position. In comparison, magnetic Bloch state $\phi^{LL}_{\bm{k}}(\bm{r})$ for Landau level defined on a torus is (up to $\bm{k}$-independent normalization)\cite{wang2021exact}\footnote{In Ref.~\cite{wang2021exact}, the magnetic field points in the negative $z$ direction, while here the magnetic field points in the positive $z$ direction.}
\begin{equation}
\phi^{LL}_{\bm{k}}(\bm{r})=\sigma(\bar{z}/l_B+il_B\bar{k}) \exp \left(i\bar{z}k\right)\exp \left(-\frac{k\bar{k}l_B^2}{2}\right) \exp \left(-\frac{z\bar{z}}{2l_B^2}\right).
\label{eq:mBlochState}\end{equation}
Therefore,
\begin{equation}
\varphi^A_{\bm{k},n}(\bm{r})=\phi^{LL}_{\bm{k}}(\bm{r})\chi_n(\bm{r})\to \psi^{0}_{\bm{k},n}(\bm{r})=\mathcal{N}_{\bm{k}} \exp \left(\frac{\gamma\bm{\nabla} ^2}{4}\right)\left [\phi^{LL}_{\bm{k}}(\bm{r})\chi_n(\bm{r})\right ]
\end{equation}
where we have set $l_B^2 = \frac{2\pi}{\Omega}$ in the magnetic Bloch state. 

To conclude, we determine the functional form of the momentum independent factor $\chi_n(\bm{r})$. To do this, we use the recursion relationship,
\begin{equation}
\exp \left(-\frac{z\bar{z}\Omega}{4\pi}\right) \chi_n(\bm{r})=\frac{(-i\sqrt{\mathcal{B}}\partial_z )^n}{\sqrt{n!}}[\exp \left(-\frac{z\bar{z}\Omega}{4\pi}\right) \chi_0(\bm{r})].
\end{equation}
This relationship can be derived from Eq.~\ref{eq:explicitphi}.
The value of $\chi_n(\bm{r})$ at arbitrary $n$ is therefore uniquely determined by the value of $\chi_0(\bm{r})$. Although we do not use the explicit expression in our calculations, we note that $\chi_0(\bm{r})$ is given by
\begin{equation}
\chi_{0}(\bm{r})=C\bar{\phi}^{LL}_{\bm{k}=\bm{0}}(\bm{r})=C \sigma(z/\sqrt{\frac{2\pi}{\Omega}})  \exp \left(-\frac{z\bar{z}\Omega}{4\pi}\right)
\end{equation}
where $C$ is a real constant.

\section{Form factors}

The form factors associated with the single particle states $\ket{\bm{k}^{(1)}}$ are
\begin{equation}
F(\bm{k}+\bm{q},\bm{k})\equiv  \braket{\lceil \bm{k}+\bm{q}\rceil^{(1)}|\exp \left(i\bm{q}\cdot\bm{r}\right)|\bm{k}^{(1)}}
\end{equation}
where $\lceil ... \rceil$ sends its argument back to the BZ. By momentum conservation, $\braket{\bm{k}_2^{(1)}|\exp \left(i\bm{q}\cdot\bm{r}\right)|\bm{k}^{(1)}_1}\propto \delta_{\bm{k}_2,\lceil \bm{k}_1+\bm{q} \rceil}$. Defining
\begin{equation}
\bm{g}_0\equiv\bm{k}+\bm{q}-\lceil \bm{k}+\bm{q} \rceil,
\end{equation}
The form factor is given by
 \begin{equation}
 \begin{split}
 F(\bm{k}+\bm{q},\bm{k}) &=A\mathcal{N}_{\lceil \bm{k}+\bm{q}\rceil}\mathcal{N}_{\bm{k}}\sum_{\bm{g}} \exp \left(-\frac{|\bm{k}+\bm{g}|^2}{4\xi^2}\right) \exp \left(-\frac{|\lceil \bm{k}+\bm{q}\rceil +\bm{g}+\bm{g}_0|^2}{4\xi^2}\right)\exp \left(-i\pi (\omega_{\bm{g}}+\frac{\bm{k}\times \bm{g}}{\Omega})\right)\\
 & \qquad \times \exp \left(i\pi \left (\omega_{\bm{g}+\bm{g}_0}+\frac{\lceil \bm{k}+\bm{q}\rceil \times (\bm{g}+\bm{g}_0)}{\Omega}\right )\right)\braket{s^{\mathcal{B}}_{\lceil \bm{k}+\bm{q} \rceil+\bm{g}+\bm{g}_0}|s^{\mathcal{B}}_{\bm{k}+\bm{g}}}\\
 &=A\mathcal{N}_{\lceil \bm{k}+\bm{q}\rceil}\mathcal{N}_{\bm{k}}\sum_{\bm{g}} \exp \left(-\frac{|\bm{k}+\bm{g}|^2}{4\xi^2}\right) \exp \left(-\frac{|\bm{k}+\bm{q}+\bm{g}|^2}{4\xi^2}\right)\exp \left(-i\pi (\omega_{\bm{g}}+\frac{\bm{k}\times \bm{g}}{\Omega_{BZ}})\right)\\
 &\qquad \times \exp \left(i\pi \left (\omega_{\bm{g}+\bm{g}_0}+\frac{\lceil \bm{k}+\bm{q}\rceil \times (\bm{g}+\bm{g}_0)}{\Omega}\right )\right)\braket{s^{\mathcal{B}}_{\bm{k}+\bm{q}+\bm{g}}|s^{\mathcal{B}}_{\bm{k}+\bm{g}}}\\
 &=A\mathcal{N}_{\lceil \bm{k}+\bm{q}\rceil}\mathcal{N}_{\bm{k}}\sum_{\bm{g}} \eta(\bm{g}_0)\exp \left(i\pi\frac{(\bm{k}+\bm{q})\times\bm{g}_0+\bm{q}\times\bm{g}}{\Omega}\right)\exp \left(-\frac{|\bm{q}|^2}{4\xi^2}\right)\exp \left(-\frac{|\bm{k}+\bm{g}|^2}{2\xi^2}\right)\\
 &\qquad \times \exp \left(-\frac{\bm{q}\cdot (\bm{k}+\bm{g})}{2\xi^2}\right)\exp \left(-\frac{\mathcal{B}}{4}(|\bm{q}|^2+2i \bm{q}\times (\bm{k}+\bm{g}) )\right)\\
 &=A\mathcal{N}_{\lceil \bm{k}+\bm{q}\rceil}\mathcal{N}_{\bm{k}} f(\bm{k})\exp \left(-\frac{\mathcal{B} |\bm{q}|^2}{4}\right) \exp \left(-\frac{ |\bm{q}|^2}{4\xi^2}\right) \eta(\bm{g}_0)\exp \left(i\pi\frac{\bm{k}\times \bm{q}+(\bm{k}+\bm{q})\times \bm{g}_0}{\Omega}\right)
 \end{split}
 \end{equation}
where $\eta(\bm{g}) \equiv \exp \left(i \pi (\omega_{\bm{g}} + 1)\right)$ is 1(-1) if $\frac{\bm{g}}{2}$ is on (not on) the reciprocal lattice. The function $f(\bm{k})$ can be written as
 \begin{equation}
 \begin{split}
 f(\bm{k})&=
 \sum_{\bm{g}}  \exp \left(-\frac{i\delta}{2} \bm{q}\times (\bm{k}+\bm{g}) \right) \exp \left(-\frac{|\bm{k}+\bm{g}|^2}{2\xi^2}\right) \exp \left(-\frac{\bm{q}\cdot (\bm{k}+\bm{g})}{2\xi^2}\right)\\
 \end{split}
 \end{equation}
To evaluate the sum over $\bm{g}$, we use Poisson summation,
\begin{equation}
f(\bm{k})=\sum_{\bm{R}}\exp \left(i\bm{k}\cdot\bm{R}\right) f(\bm{R})
\end{equation}
\begin{equation}
\begin{split}
f(\bm{R})&=\frac{1}{\Omega}\int d\bm{k}  \exp \left(-\frac{i\delta}{2} \bm{q}\times \bm{k} \right) \exp \left(-\frac{|\bm{k}|^2}{2\xi^2}\right) \exp \left(-\frac{\bm{q}\cdot \bm{k}}{2\xi^2}\right) \exp \left(-i\bm{k}\cdot \bm{R}\right)\\
&=\frac{2\pi\xi^2}{\Omega} \exp \left(-\frac{|\bm{R}|^2\xi^2}{2}\right)\exp \left(\frac{i\bm{q}\cdot \bm{R}}{2}\right)\exp \left(\frac{\xi^2\delta (\bm{R}\times \bm{q})}{2}\right)\exp \left(\frac{|\bm{q}|^2(1-\delta^2\xi^4)}{8\xi^2}\right)
\label{eq:Resummedf}\end{split}
\end{equation}
where $\bm{R}$ are discrete points on the real space lattice that is dual to $\bm{g}$, i.e., $\bm{R}\cdot \bm{g} \in 2\pi \mathbb{Z}$ for all $\bm{R}$ and $\bm{g}$. In Eq.~\ref{eq:Resummedf}, we find that $f(\bm{R})$ is exponentially suppressed by a factor of $\exp \left(-\frac{|\bm{R}|^2\xi^2}{2}\right)$. Since $\bm{R}$ live on the dual lattice, spanned by $\bm{g}_1$ and $\bm{g}_2$, $\exp \left(-\frac{|\bm{R}|^2\xi^2}{2}\right)\leq \epsilon$ for $\bm{R}\neq 0$, where
\begin{equation}
    \epsilon \equiv \exp \left(-\frac{4\pi^2\xi^2}{\sqrt{3}\Omega}\right)
\label{eq:epsilonDef}\end{equation}
If we are considering a state where $\epsilon$ is small, all $f(\bm{R})$ with $\bm{R}\neq 0$ will be highly suppressed and can be ignored. In this limit, $f(\bm{k})$ is constant, independent of momentum. As we shall discuss, many of the assumptions we make assume that $\epsilon$ is small. The value of $\epsilon$ in the variationally optimized states is shown in Sec.~\ref{sup:ErrAnalysis}. 

We now consider the normalization factors,
\begin{equation}
    \frac{1}{\mathcal{N}_{\bm{k}}^2} = A\sum_{\bm{g}}  \exp \left(-\frac{|\bm{k}+\bm{g}|^2}{2\xi^2}\right)
\end{equation}
Using Poisson summation to evaluate the sum over $\bm{g}$, we find that
\begin{equation}
\frac{1}{\mathcal{N}_{\bm{k}}^2}= A\sum_{\bm{g}}\exp \left(-\frac{|\bm{k}+\bm{g}|^2}{2\xi^2}\right)=\frac{2\pi \xi^2 A}{\Omega}+O(\epsilon)
\label{eq:NormalizationDef}\end{equation}
where $A$ is the system area. Again,  assuming that $\epsilon$ is sufficiently small,  $\mathcal{N}_{\bm{k}}$ can be treated as being independent of momentum.

Using the momentum independent forms of $f(\bm{k})$ and $\mathcal{N}_{\bm{k}}$, the form factors are 
\begin{equation}
F(\bm{k+q},\bm{k})=\exp \left(-\frac{|\bm{q}|^2}{8\xi^2_1}\right)\eta(\bm{g}_0)
\exp[-\frac{\pi}{2\Omega}(|\bm{q}|^2-2i \bm{k}\times \bm{q}-2i(\bm{k}+\bm{q})\times \bm{g}_0)] ,
\label{eq:FormFactorSimp}\end{equation}
Where we have defined 
\begin{equation}
\xi^2_1\equiv \frac{\xi^2}{(1+\delta\xi^2)^2}.
\label{eq:xi1Def}\end{equation}
These form factors are very similar to the form factors for magnetic Bloch states in the lowest Landau level (LLL) with magnetic field pointing in the positive $\hat{z}$ direction \cite{tan2024parent},
\begin{equation}
\begin{split}
F_{LLL}(\bm{k}+\bm{q},\bm{k})&\equiv \braket{\lceil \bm{k}+\bm{q} \rceil^{LL}|\exp \left( i\bm{q}\cdot \bm{r} \right )|\bm{k}^{LL}}=\eta(\bm{g}_0)\exp \left(-i\frac{l_B^2}{2} (\bm{q}\times \bm{k}+\bm{g}_0\times \lceil \bm{k}+\bm{q} \rceil)\right)
\exp \left(-\frac{l_B^2|\bm{q}|^2}{4}\right),
\end{split}
\end{equation}
where $\phi^{LL}_{\bm{k}}(\bm{r})=\braket{\bm{r}|\bm{k}^{LL}}$ are the magnetic Bloch states, $\bm{k}$ is a momentum in the magnetic BZ, and $l_B$ is the magnetic length. If we identify $l_B^2 = \frac{2\pi}{\Omega}$, then we arrive at the simple expression 
\begin{equation}
F(\bm{k}+\bm{q},\bm{k})=\exp \left(-\frac{|\bm{q}|^2}{8\xi_1^2}\right)F_{LLL}(\bm{k}+\bm{q},\bm{k}),
\label{eq:FormFactorsDef}\end{equation}
which is valid in the limit when $\xi^2/\Omega$ is significantly large. 

\section{Energy of the IAHC and FAHC ansatz wavefunctions}
In this section we determine the energies of the IAHC and FAHCs. Again, we consider the limit where $\epsilon$ is small in this section. 
\subsection{Kinetic Energy}
To evaluate the kinetic energy (see Sec.~\ref{sup:ICB}) of the IAHC and FAHCs, we first expand the single particle creation and annihilation operator in terms of the Bloch operators,
\begin{equation}
c_{n,\bm{r}}=\sum_{\bm{k}} \psi^{(1)}_{\bm{k},n}(\bm{r}) c^{(1) }_{\bm{k}},\quad c^{\dagger}_{n,\bm{r}}=\sum_{\bm{k}} \psi^{(1)*}_{\bm{k},n}(\bm{r}) c^{(1) \dagger}_{\bm{k}},
\end{equation}
where $c^{(1) \dagger}_{\bm{k}}$ is the operator that creates an electron in the Bloch state $\ket{\bm{k}^{(1)}}$. The kinetic energy per particle in the anomalous Hall crystal (AHC) is then
\begin{equation}
\begin{split}
\frac{E^{\text{kin}}}{N}&=\sum_{\bm{k}_1,\bm{k}_2}-\frac{1}{2m_0 N}\braket{\bm{k}^{(1)}_1|\nabla^2|\bm{k}^{(1)}_2}\langle c^{(1) \dagger}_{\bm{k}_1} c^{(1) }_{\bm{k}_2}\rangle_{\text{AHC}} \\
&=\sum_{\bm{k},\bm{g}} \mathcal{N}^2_{\bm{k}}A\frac{|\bm{k}+\bm{g}|^2}{2m_0 N}\exp \left(-\frac{(\bm{k}+\bm{g})^2}{2\xi^2}\right)\langle  c^{(1) \dagger}_{\bm{k}} c^{(1) }_{\bm{k}}\rangle_{\text{AHC}}
\end{split}
\end{equation}

Making use of the Poission summation formula
\begin{equation}
\sum_{\bm{g}}\frac{|\bm{k}+\bm{g}|^2}{2m_0}\exp \left(-\frac{(\bm{k}+\bm{g})^2}{2\xi^2}\right)=\sum_{\bm{R}}\exp \left(i\bm{k}\cdot \bm{R}\right)\frac{\pi \xi^4(2-|\bm{R}|^2\xi^2)}{m_0\Omega}\exp \left(-\frac{\xi^2|\bm{R}|^2}{2}\right)
\end{equation}
If we ignore terms that are supressed in powers of $\epsilon$, the kinetic energy per particle is
\begin{equation}
\frac{E^{\text{kin}}}{N} = \sum_{\bm{k}}\frac{\xi^2}{m_0 N} \langle c^{(1) \dagger}_{\bm{k}}c^{(1) }_{\bm{k}} \rangle_{\text{AHC}} = \frac{\xi^2}{m_0},
\end{equation}
where we have used that $\sum_{\bm{k}}\braket{c^{(1) \dagger}_{\bm{k}}c^{(1) }_{\bm{k}}}_\text{AHC}=N$ for both the IAHC and FAHC. 

\subsection{Interaction Energy}
In this work, we consider the variational energies with respect to the density-density Coulomb interaction
\begin{equation}
V(\bm{r})=\frac{V_{\text{C}}}{2\pi |\bm{r}|} \quad V(\bm{q})=\int d\bm{r} V(\bm{r}) \exp \left(-i\bm{q}\cdot \bm{r}\right)=\frac{V_{\text{C}}}{ |\bm{q}|}.
\label{eq:CInteraction}\end{equation}
In second quantized notation, the interaction term in the Hamiltonian is
\begin{equation}
    \hat{\mathcal{H}}^{\text{int}} =\int d\bm{r}d\bm{r}'\sum_{n,n'}\frac{1}{2}V(\bm{r}-\bm{r}') c^\dagger_{n,\bm{r}}c^\dagger_{n',\bm{r}'}c_{n',\bm{r}'}c_{n,\bm{r}}=\int d\bm{r}d\bm{r}'\frac{1}{2}V(\bm{r}-\bm{r}') :\hat{\rho}(\bm{r})\hat{\rho}(\bm{r}'):
\end{equation}
where $\hat{\rho}(\bm{r})$ is the density operator, $:..:$ stands for normal ordering. $V_{\text{C}}$ is the interaction strength. To find the interaction energy, we will also subtract off a divergent charging energy term, associated with the long range nature of the Coulomb. This will be discussed more in the following subsections. 

\subsubsection{Second-Quantized Representation of the Ansatz Wavefunction}
To evaluate the interaction energy, we use a second-quantized representation of the ansatz wavefunctions. Previously, we showed that
\begin{equation}
\psi^{(1)}_{\bm{k},n}(\bm{r})=\mathcal{N}_{\bm{k}} \exp \left(\frac{\gamma\bm{\nabla} ^2}{4}\right)[\phi^{LL}_{\bm{k}}(\bm{r})\chi_n(\bm{r})]
\end{equation}
Explicitly, we consider the situation where $\psi^{(1)}_{\bm{k},n}$ is defined on the torus, as is $\phi^{LL}_{\bm{k}}$. The (unnormalized) many-body wavefunction for the integer anomalous Hall crystal (IAHC) is given by the Slater determinant of all $\psi^{(1)}_{\bm{k},n}$,
\begin{equation}\begin{split}
\Psi^{\mathrm{IAHC}}_{\{n\}}(\{\bm{r}\}) \! &= \exp \left(\frac{\gamma}{4} \sum_{i} \bm{\nabla}^2_i \right)
\begin{vmatrix}
\phi^{LL}_{\bm{k}_1}(\bm{r}_1)& \phi^{LL}_{\bm{k}_1}(\bm{r}_2)..&\phi^{LL}_{\bm{k}_1}(\bm{r}_N)\\
\phi^{LL}_{\bm{k}_2}(\bm{r}_1)& ..&\phi^{LL}_{\bm{k}_2}(\bm{r}_N)\\
...\\
\phi^{LL}_{\bm{k}_N}(\bm{r}_1)& \phi^{LL}_{\bm{k}_N}(\bm{r}_2)..&\phi^{LL}_{\bm{k}_N}(\bm{r}_N)
\end{vmatrix}
\prod_i \chi_{n_i} (\bm{r}_i)  \\\\  &=\exp \left( \frac{\gamma}{4} \sum_{i}\bm{\nabla}^2_i\right) \left [\Phi^{\mathrm{IQH}}(\{\bm{r} \}) \prod_i \chi_{n_i} (\bm{r}_i)\right],
\end{split}\end{equation}
where $\Phi^{\mathrm{IQH}}$ is the many-body wavefunction of the fully filled LLL. If we were to instead consider a disk geometry, $\Phi^{\mathrm{IQH}}$ would be the fully filled LLL wavefunction on a disk,
\begin{equation}
    \Phi^{\mathrm{IQH}}(\{\bm{r}\}) = \prod_{i < j} (\bar{z}_i-\bar{z}_j) \exp \left (- \sum_i |z_i|^2/2l^2_B\right).
\end{equation}
Note that, as a fermionic many-body wavefunction, $\Psi^{\text{IAHC}}$ is antisymmetric under exchange of $(\bm{r}_i, n_i)\leftrightarrow (\bm{r}_j,n_j)$. 

We now turn to the ansatz wavefunction for the fractional anomalous Hall crystal (FAHC),
\begin{equation}
\Psi^{\mathrm{FAHC}}_{\{n\}}(\{\bm{r}\}) \! = \exp \left(\! \frac{\gamma}{4} \sum_{i}\bm{\nabla}^2_i\!\right) \left[ \Phi^{\mathrm{FQH}}(\{\bm{r}\}) \prod_i \chi_{n_i}(\bm{r}_i)\right ].
\label{eq:FAHCwfSN}\end{equation}
In the torus geometry, $\Phi^{\mathrm{FQH}}(\{\bm{r} \})$ can be decomposed into linear combination of Slater determinants states formed out of the magnetic Bloch state $\phi^{LL}_{\bm{k}}(\bm{r})=\braket{\bm{r}|\bm{k}^{LL}}$ in the LLL,
\begin{equation}
\ket{\Phi^{\text{FQH}}}=\sum_{\{\bm{k}\}} C_{\{\bm{k}\}} \ket{\{ \bm{k}^{LL}\}} \equiv \sum_{\bm{k}_1,..\bm{k}_N} 
C_{\{\bm{k}\}} \ket{\bm{k}^{LL}_1...\bm{k}^{LL}_N}, 
\end{equation}
where $ C_{\{\bm{k}\}}$ are the complex coefficients of different single particle states. Using this, the FAHC wavefunction is
\begin{equation}
\begin{split}
\Psi^{\mathrm{FAHC}}_{\{n\}}(\{\bm{r}\}) \!& =  \sum_{\{\bm{k}\}}  C_{\{\bm{k}\}}
\exp \left(\! \frac{\gamma}{4} \sum_{i}\bm{\nabla}^2_i\!\right) \begin{vmatrix}
\phi^{LL}_{\bm{k}_1}(\bm{r}_1)& \phi^{LL}_{\bm{k}_1}(\bm{r}_2)..&\phi^{LL}_{\bm{k}_1}(\bm{r}_N)\\
\phi^{LL}_{\bm{k}_2}(\bm{r}_1)& ..&\phi^{LL}_{\bm{k}_2}(\bm{r}_N)\\
...\\
\phi^{LL}_{\bm{k}_N}(\bm{r}_1)& \phi^{LL}_{\bm{k}_N}(\bm{r}_2)..&\phi^{LL}_{\bm{k}_N}(\bm{r}_N)
\end{vmatrix}\prod_{i}
\chi_{n_i}(\bm{r}_i)\\\\
&=\sum_{\{\bm{k}\}} \frac{C_{\{\bm{k}\}}}{\mathcal{N}_{\{\bm{k}\}}}\begin{vmatrix}
\psi^{(1)}_{\bm{k}_1,n_1}(\bm{r}_1)& \psi^{(1)}_{\bm{k}_1,n_2}(\bm{r}_2)..&\psi^{(1)}_{\bm{k}_1,n_N}(\bm{r}_N)\\
..&..&..\\
\psi^{(1)}_{\bm{k}_N,n_1}(\bm{r}_1)& \psi^{(1)}_{\bm{k}_N,n_2}(\bm{r}_2)..&\psi^{(1)}_{\bm{k}_N,n_N}(\bm{r}_N)
\end{vmatrix}
\end{split}
\end{equation}
where we defined $\mathcal{N}_{\{\bm{k}\}}\equiv \prod_{\bm{k}\in \{\bm{k}\}}\mathcal{N}_{\bm{k}}$. We now make use of the fact that $\mathcal{N}_{\bm{k}}$ is $\bm{k}$-independent to leading order in $\epsilon$. If we ignore the $\bm{k}$-dependence of $\mathcal{N}_{\bm{k}}$ the FAHC wavefunction in the second-quantized notation is simply (up to normalization)
\begin{equation}\label{eq:exactWF}
\ket{\Psi^{\text{FAHC}}}= \sum_{\{\bm{k}\}} C_{\{\bm{k}\}} \ket{\{ \bm{k}^{(1)}\}}
\end{equation}
To conclude, our FAHC wavefunction, expressed as linear combination of Slater determinants of Bloch states, has the same coefficients as the FQH wavefunction expressed as linear combination of Slater determinants of magnetic Bloch states when $\epsilon$ is sufficiently small.


\subsubsection{Pair Correlation Function}
The interaction energy per particle in the FAHC/IAHC can be written as
\begin{equation}
\begin{split}
\frac{E^{\text{int}}}{N}&=\frac{(N-1)}{2}\sum_{\{ n_i \}}\int \prod_i d\bm{r}_i  V(\bm{r}_1-\bm{r}_2) |\Psi^{\text{AHC}}_{\{ n_i\}}(\bm{r}_1,..,\bm{r}_N)|^2\\
&=\frac{(N-1)}{2}\sum_{\{n_i\}} \int  d\bm{r} V(\bm{r}) [\int \prod_{i>1}d\bm{r}_i |\Psi^{\text{AHC}}_{\{ n_i\}}(\bm{r}_2+\bm{r},..,\bm{r}_N)|^2]\\
&\equiv \frac{\rho_0}{2} \int d\bm{r}g^{\text{AHC}}(\bm{r}) V(\bm{r}),
\end{split}
\end{equation}
where $\Psi^{\text{AHC}}$ is the appropriate IAHC/FAHC many-body wavefunction, $\rho_0$ is the average particle density, and we have defined the pair correlation function $g(\bm{r})$ (also known as radial distribution function and two-point function\cite{PhysRevB.33.2481,PhysRevB.30.558})
\begin{equation}
g^{\text{AHC}}(\bm{r})\equiv\frac{N-1}{\rho_0} \sum_{\{n_i\}} \int \prod_{i>1}d\bm{r}_i |\Psi^{\text{AHC}}_{\{ n_i\}}(\bm{r}_2+\bm{r},..,\bm{r}_N)|^2.
\end{equation}
In second-quantized notation, $g(\bm{r})$ is the expectation value of the two-body operator
\begin{equation}
g^{\text{AHC}}(\bm{r}) =\int \frac{1}{A} g^{\text{AHC}}(\bm{r}'+\bm{r},\bm{r}') d\bm{r}' \equiv \frac{1}{\rho_0 N}\sum_{n,n'} \int d\bm{r}^\prime \langle   c^{\dagger}_{n,\bm{r}'+\bm{r}} c^{\dagger}_{n',\bm{r}'}c_{n',\bm{r}'}c_{n,\bm{r}'+\bm{r}}\rangle_{\text{AHC}}
\end{equation}
Note that $g(\bm{r})\to 1$ as $|\bm{r}|\to \infty$, and this contribution would be divergent since Coulomb interaction is long-range, i.e. $V(\bm{q}=0)$ diverges. This part should be removed from the definition of the interaction energy such that
\begin{equation}
\begin{split}
\frac{E^{\text{int}}}{N}
&= \frac{\rho_0}{2} \int d\bm{r}(g^{\text{AHC}}(\bm{r})-1) V(\bm{r}) \equiv \frac{\rho_0}{2} \int d\bm{r}g^{\text{AHC}}_1(\bm{r})V(\bm{r}).
\end{split}
\end{equation}
This is the definition of the interaction energy used in the main text. 

We now proceed to show that the pair correlation functions of the anomalous Hall crystals can be expressed in terms of the pair correlation functions of the corresponding quantum Hall states. 
The IAHC/FAHC pair correlation function is
\begin{equation}
g^{\text{AHC}}(\bm{r})=\frac{1}{\rho_0 N}\sum_{n,n'}\sum_{\bm{k}_1...\bm{k}_4}\int d\bm{r}'\langle c^{(1) \dagger}_{\bm{k}_1}c^{(1) \dagger}_{\bm{k}_2}c^{(1) }_{\bm{k}_4}c^{(1) }_{\bm{k}_3} \rangle_{\text{AHC}} \times \left [ \psi^{(1)}_{\bm{k}_1,n}(\bm{r}'+\bm{r})\psi^{(1)*}_{\bm{k}_3,n}(\bm{r}'+\bm{r})  \psi^{(1)}_{\bm{k}_2,n'}(\bm{r}')\psi^{(1)*}_{\bm{k}_4,n'}(\bm{r}')\right ]
\end{equation}
where $\langle ... \rangle_{\text{AHC}}$ is evaluated with respect to the appropriate AHC trial wavefunction. Since the IAHC and FAHC states conserve crystal momentum,
\begin{equation}
\begin{split}
g^{\text{AHC}}(\bm{r})=\frac{1}{\rho_0 N}\sum_{n,n'}\sum_{\bm{k}_1,\bm{k}_2 \in BZ ,\bm{q}_1}\int d\bm{r}'&\langle c^{(1) \dagger}_{\bm{k}_1}c^{(1) \dagger}_{\bm{k}_2}c^{(1)}_{\lceil \bm{k}_2+\bm{q}_1 \rceil }c^{(1)}_{\lceil \bm{k}_1-\bm{q}_1 \rceil }\rangle_{\text{AHC}}\\ &\times \left [\psi^{(1)}_{\bm{k}_1,n}(\bm{r}'+\bm{r})\psi^{(1)*}_{\lceil \bm{k}_1-\bm{q}_1 \rceil,n}(\bm{r}'+\bm{r})  \psi^{(1)}_{\bm{k}_2,n'}(\bm{r}')\psi^{(1)*}_{\lceil \bm{k}_2+\bm{q}_1 \rceil,n'}(\bm{r}')\right ]
\end{split}
\end{equation}
The Fourier transform of $g^{\text{AHC}}(\bm{r})$ is
\begin{equation}
\begin{split}
g^{\text{AHC}}(\bm{q})&=\int d\bm{r} g^{\text{AHC}}(\bm{r}) \exp \left(-i\bm{r}\cdot\bm{q}\right)\\
&=\frac{1}{ \rho_0 N}\sum_{\bm{k}_1,\bm{k}_2 \in BZ , \bm{q}_1 }\delta_{\lceil \bm{q}_1-\bm{q}\rceil,\bm{0}}\langle c^{(1) \dagger}_{\bm{k}_1}c^{(1) \dagger}_{\bm{k}_2}c^{(1)}_{\lceil \bm{k}_2+\bm{q}_1 \rceil }c^{(1)}_{\lceil \bm{k}_1-\bm{q}_1 \rceil } \rangle_{\text{AHC}} F(\bm{k}_2+\bm{q},\bm{k}_2)  F(\bm{k}_1-\bm{q},\bm{k}_1)
\end{split}
\end{equation}
We now compare $g^{\text{AHC}}(\bm{q})$ to the pair correlation functions of the corresponding IQH/FQH wavefunction,
\begin{equation}\label{eq:numericalPair}
g^{\text{QH}}(\bm{q})=\frac{1}{\rho_0 N}\sum_{\bm{k}_1,\bm{k}_2 \in MBZ , \bm{q}_1 }\delta_{\lceil \bm{q}_1-\bm{q}\rceil,\bm{0}}\braket{c^{LL\dagger}_{\bm{k}_1}c^{LL\dagger}_{\bm{k}_2}c^{LL}_{\lceil \bm{k}_2+\bm{q}_1 \rceil }c^{LL}_{\lceil \bm{k}_1-\bm{q}_1 \rceil }}_{\text{QH}} F_{\text{LLL}}(\bm{k}_2+\bm{q},\bm{k}_2)  F_{\text{LLL}}(\bm{k}_1-\bm{q},\bm{k}_1)
\end{equation}
Here, $c^{M\dagger}_{\bm{k}}$ creates an electron in the magnetic Bloch state $\ket{\bm{k}^{LL}}$. The expectation value $\langle ... \rangle_{\text{QH}}$ is evaluated with respect to the appropriate QH wavefunction. Since both the  IAHC and IQH wavefunctions are simple Slater determinants, $\braket{c^{LL\dagger}_{\bm{k}_1}c^{LL\dagger}_{\bm{k}_2}c^{LL}_{\lceil \bm{k}_2+\bm{q}_1 \rceil }c^{LL}_{\lceil \bm{k}_1-\bm{q}_1 \rceil }}_{\text{IQH}}=\langle c^{(1) \dagger}_{\bm{k}_1}c^{(1) \dagger}_{\bm{k}_2}c^{(1)}_{\lceil \bm{k}_2+\bm{q}_1 \rceil }c^{(1)}_{\lceil \bm{k}_1-\bm{q}_1 \rceil }\rangle_{\text{IAHC}}$. Similarly, from Eq.\ref{eq:exactWF}, $\braket{c^{LL\dagger}_{\bm{k}_1}c^{LL\dagger}_{\bm{k}_2}c^{LL}_{\lceil \bm{k}_2+\bm{q}_1 \rceil }c^{LL}_{\lceil \bm{k}_1-\bm{q}_1 \rceil }}_{\text{FQH}}=\langle c^{(1) \dagger}_{\bm{k}_1}c^{(1) \dagger}_{\bm{k}_2}c^{(1)}_{\lceil \bm{k}_2+\bm{q}_1 \rceil }c^{(1)}_{\lceil \bm{k}_1-\bm{q}_1 \rceil }\rangle_{\text{FAHC}}$. 

Using this and the previously established connection between the IAHC and LLL form factors (Eq.~\ref{eq:FormFactorsDef}), we arrive at the  conclusion that
\begin{equation}
 g^{\text{AHC}}(\bm{q})=\exp \left(-\frac{|\bm{q}|^2}{4\xi_1^2}\right) g^{\text{QH}}(\bm{q})
\end{equation}
We again stress that we have identified $l_B^2 = \frac{2\pi}{\Omega}$, and assumed that $\epsilon$ is sufficiently small. 

\subsubsection{Pair Correlation Function of the Integer Quantum Hall State}
The pair correlation function for the fully filled LLL is analytically known to be\cite{PhysRevB.33.2481,Jain_2007}
\begin{equation}
g^{\text{IQH}}(\bm{r})=1-\exp \left(-\frac{|\bm{r}|^2}{2l_B^2}\right),\qquad g^{\text{IQH}}_1(\bm{q})=\int d\bm{r}(g^{\text{IQH}}(\bm{r})-1) \exp \left(-i\bm{q}\cdot\bm{r}\right)=-2\pi l_B^2\exp \left(-\frac{l_B^2 |\bm{q}|^2}{2}\right).
\end{equation}
The $-1$ we inserted when computing $g_1(\bm{q})$ is needed to remove the divergent charging energy of the system. Using the analytic form of $g^{\text{IQH}}$, the interaction energy per particle for IAHC can be directly evaluated as
\begin{equation}
\frac{E^{\text{int}}_{\text{IAHC}}}{N}= \frac{1}{2}\rho_0 \int d\bm{q}\frac{1}{4\pi^2}g^{\text{IAHC}}_1(\bm{q})V(\bm{q})=-\frac{V_c}{4\sqrt{\pi}}\frac{1}{\sqrt{1/\xi_1^2+\frac{4\pi}{\Omega}}}.
\label{eq:IAHCIntEnergy}\end{equation}

\subsubsection{Pair Correlation Functions of the Fractional Quantum Hall States}

The pair-correlation function for FQH state is not known analytically. However, they can be computed and fitted numerically.  We use the Eq. \ref{eq:numericalPair} for numerical computation of $g^{\text{FQH}}(\bm{q})$.
The FQH wavefunction is obtained numerically via Exact Diagonalization(ED) on a torus. Notice that for FQH wavefunctions at filling factor $\nu$, $\rho_0=\frac{\nu}{2\pi l_B^2}$.  

\begin{figure}[t!]
    \centering
    \includegraphics[width=0.98\linewidth]{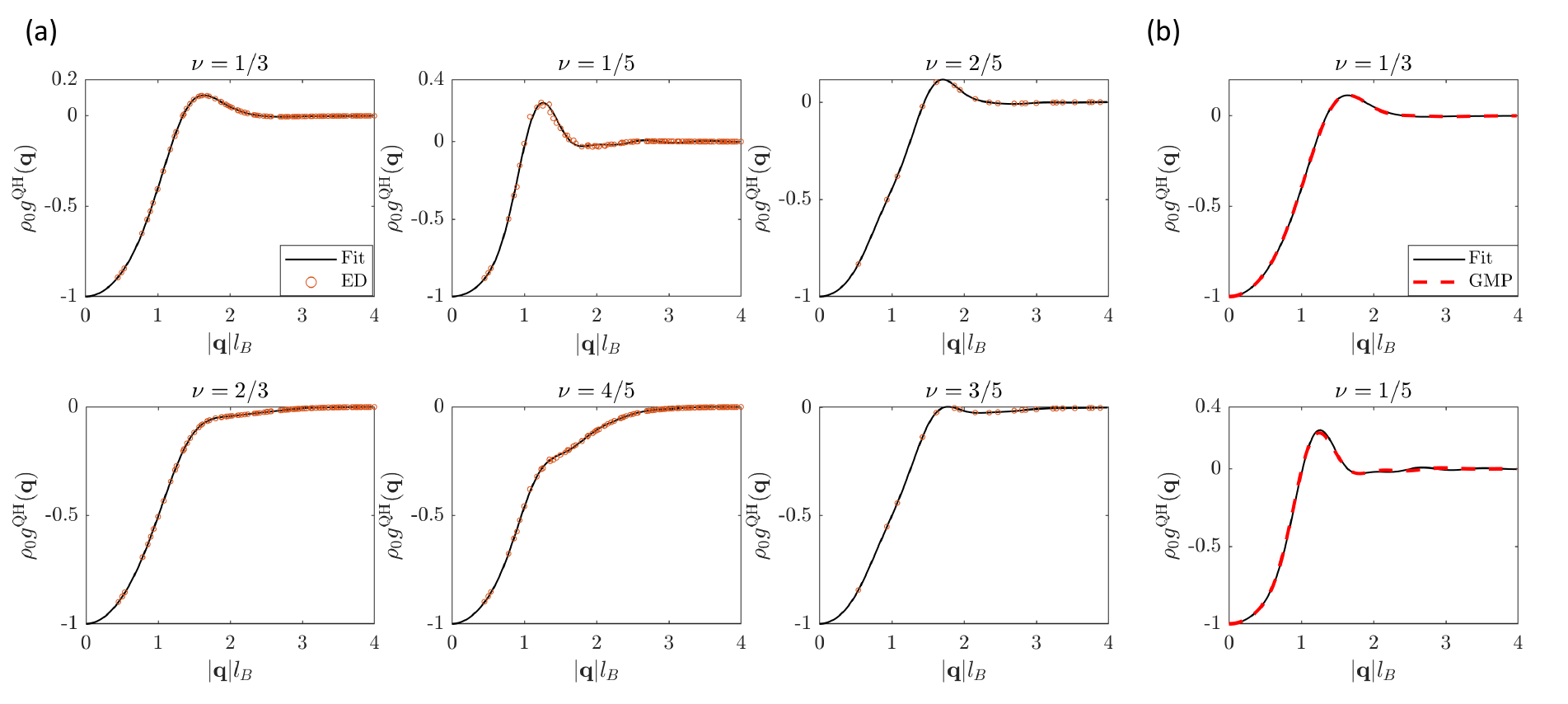}
    \caption{(a)Pair-correlation function of fractional quantum Hall states at filling factor $\nu$. $q=\bm{0}$ is excluded from the numerical computation. The red circles are numerical data obtained from Exact Diagonalization, while black lines are our fitting results. We use 5-by-6 30-site cluster for $\nu=\frac{1}{3}$/$\frac{2}{3}$/$\frac{1}{5}$/$\frac{4}{5}$, and 5-by-5 25-site cluster for $\nu=\frac{2}{5}$/ $\frac{3}{5}$. To obtain the fractional quantum Hall state, we use pseudopotential $V(\bm{q})=-|\bm{q}|^2$ for $\nu=\frac{1}{3}$/$\frac{2}{3}$, $V(\bm{q})=-|\bm{q}|^6$ for $\nu=\frac{1}{5}$/$\frac{4}{5}$, and Coulomb interaction $V(\bm{q})=\frac{1}{|\bm{q}|}$ for $\nu=\frac{3}{5}$/$\frac{2}{5}$. (b) Comparison of our fitting result (black solid line) with that presented in \cite{PhysRevB.33.2481} (GMP, red dashed line).  }
    \label{SPfig_pairfit}
\end{figure}

It was proposed in Ref.\cite{PhysRevB.30.558} that FQH pair correlation function can be fitted using
\begin{equation}
g^{\text{FQH}}(\bm{r})=1-\exp \left(-\frac{|\bm{r}|^2}{2l_B^2}\right)+\sum_{m\in odd}\frac{2}{m!}(\frac{|\bm{r}|^2}{4l_B^2})^m c_m \exp \left(-\frac{|\bm{r}|^2}{4l_B^2}\right),
\end{equation}
or, in momentum space,
\begin{equation}
g^{\text{FQH}}(\bm{q})=-2\pi l_B^2\exp \left(-\frac{l_B^2 |\bm{q}|^2}{2}\right)\\
+\sum_{m\in odd}8\pi l_B^2c_m L_m(l_B^2 |\bm{q}|^2)\exp \left(-l_B^2 |\bm{q}|^2)+4\pi^2\delta(\bm{q}\right),
\end{equation}
where  $c_m$ are the fitting coefficients, and $L_m$ are the Lageurre polynomials. To avoid the divergence, associated with the $\delta(\bm{q})$, we manually exclude $\bm{q}=0$ from our calculation. The following sum rules are also imposed when fitting a FQH state at filling $\nu$\cite{PhysRevB.30.558}.
\begin{equation}
\sum_{m\in odd} c_m=(1-1/\nu)/4,\quad \sum_{m\in odd}(m+1)c_m=(1-1/\nu)/8 ,\quad \sum_{m\in odd}(m+1)(m+2)c_m=(1-1/\nu)^2/8.
\end{equation}
The fitting coefficients for $\nu=\frac{1}{3}$ and $\nu=\frac{1}{5}$ Laughlin state are given in Ref.\cite{PhysRevB.30.558} (and equivalently  $\nu=\frac{2}{3}$ and $\nu=\frac{4}{5}$ state, since they are related to the Laughlin state by particle-hole conjugation). We use Coulomb interaction to obtain the ground state at $\nu=\frac{2}{5}$ and $\nu=\frac{3}{5}$ Jain states. The fitting result is listed in Table.\ref{tb:fitresult} and plots are shown in Fig~\ref{SPfig_pairfit}.

\begin{center}
\begin{table}
\begin{tabular}{ |c|c|c|c|c|c|c|c|c| } 
\hline
\quad & $\nu=\frac{1}{3}$ & $\frac{2}{3}$ & $\frac{1}{5}$ &$\frac{4}{5}$ &$\frac{2}{5}$ & $\frac{3}{5}$ & $\text{GMP}, \frac{1}{3}$ &$\text{GMP}, \frac{1}{5}$ \\
\hline
$c_1$ & -1.0013 & -0.2509 &-0.9930 &-0.0627 &-0.7934 &-0.3530 &-1.0&-1.0 \\ 
\hline
$c_3$ & 0.4748 & 0.1199 &-1.1447 &-0.0700 &0.5022 &0.2244 &0.51053&-1.0\\
\hline
$c_5$ & 0.1213 & 0.0250 &1.3311 &0.0765 &0.0151 &0.0018 &-0.02056&0.6765\\
\hline
$c_7$ & 0.0642 & 0.0328 &-2.1470 &-0.1132 &0.1961 &0.1036 &0.31003&0.3130\\
\hline
$c_9$ & -0.3357 & -0.1219 &6.8035 &0.3777 &-0.9976 &-0.4831 &-0.49050&-0.1055\\
\hline
$c_{11}$ & 0.4315 & 0.1701 &-11.3828 &-0.6336 &1.5539 &0.7566 &0.20102&0.8910\\
\hline
$c_{13}$ & -0.5221 & -0.1988 &13.1108 &0.7340 &-1.8382 &-0.8898 &-0.00904&-0.3750\\
\hline
$c_{15}$ & 0.3545 & 0.1347 &-9.3319 &-0.5257 &1.4846 &0.7086 &-0.00148&-0.7750\\
\hline
$c_{17}$ & -0.0874 & -0.0359 &2.7540 &0.1546 &-0.4978 &-0.2357 &0.0&0.37\\
\hline
$c_{19}$ & 0.0 & 0.0 &0.0 &0.0 &0.0 &0.0 &0.0012&0.01\\
\hline
$c_{21}$ & 0.0 & 0.0 &0.0 &0.0 &0.0 &0.0 &0.0006&-0.005\\
\hline
$c_{23}$ & 0.0 & 0.0 &0.0 &0.0 &0.0 &0.0 &-0.0018&-0.0\\
\hline
$c_{25}$ & 0.0 & 0.0 &0.0 &0.0 &0.0 &0.0 &0.0&-0.1\\
\hline
$c_{27}$ & 0.0 & 0.0 &0.0 &0.0 &0.0 &0.0 &0.0&0.1\\
\hline
\end{tabular}
\caption{Fitting coefficients for the pair correlation function $\bm{g}^{\text{QH}}(\bm{q})$. The rightmost two columns are those presented in Ref.\cite{PhysRevB.33.2481}, which we include here for comparison. We truncate our fitting at $c_{17}$. Allowing more fitting coefficients does not change the fitting curve in an observable way.}\label{tb:fitresult}
\end{table}
\end{center}

\section{Wannier approach for the energy of WC and IAHC}
In this section, we use Wannier functions to calculate the energy of the WC and IAHC. We note that the energy of the IAHC obtained in this method is consistent with the energy obtained by using the pair correlation function discussed in the previous section.

\subsection{Wigner Crystal}
The ansatz wavefunction for WC is a Slater determinant of the following single particle states,
\begin{equation}
\braket{n,\bm{r}|\bm{k}^{(0)}}=\psi^{(0)}_{\bm{k},n}(\bm{r})=\sum_{\bm{g}}\mathcal{N}_{\bm{k}}\exp \left(-\frac{|\bm{k}+\bm{g}|^2}{4\xi^2}\right)w_{\bm{k}+\bm{g},n}(\bm{r})
\end{equation}
Since the $\psi^{(0)}_{\bm{k},n}$ wavefunction has zero Chern number, it can be Wannierized. The corresponding Wannier wavefunction $\varphi_{\bm{R},n}$ are defined via
\begin{equation}
\psi^{(0)}_{\bm{k},n}(\bm{r})=\sum_{\bm{R}}\frac{1}{\sqrt{N}} \exp \left(i\bm{k}\cdot \bm{R})  \varphi_{\bm{R},n}(\bm{r}\right)
\end{equation}
where $N$ is the number of unit cells (which is also equal to the number of electrons in this case), and $\bm{R}$ is summed over the real space lattice vectors. Using translation symmetry, we can define $\varphi_{\bm{R},n}(\bm{r})\equiv \varphi_n(\bm{r}-\bm{R})$, where, explicitly,
\begin{equation}
\begin{split}
\varphi_n(\bm{r})&=\frac{\sqrt{N}}{\Omega} \int  d \bm{q} \mathcal{N}_{\lceil \bm{q} \rceil} \exp \left(-\frac{|\bm{q}|^2}{4\xi^2}\right)  w_{\bm{q},n}
\end{split}
\end{equation}
The charge density $n_0(\bm{r})$ associated with the $\varphi_n(\bm{r})$ is,
\begin{equation}
\begin{split}
n_0(\bm{r})&=\sum_{n}|\varphi_n(\bm{r})|^2\\
&=\frac{N}{\Omega^2}\int d\bm{q}_1 d\bm{q}_2   \mathcal{N}_{\lceil \bm{q}_1 \rceil}\mathcal{N}_{\lceil \bm{q}_2 \rceil} \exp \left(-(\bm{q}_1^2+\bm{q}_2^2)/4\xi^2\right)  \exp \left(i(\bm{q}_1-\bm{q}_2)\cdot\bm{r}\right)\exp \left(-\frac{\mathcal{B}}{4}(|\bm{q}_1 - \bm{q}_2|^2 + 2i \bm{q}_1 \times \bm{q}_2 )\right)\\
&=\frac{2}{\pi}\frac{\xi^2}{(1+\mathcal{B}\xi^2)^2}\exp \left(-\frac{2|\bm{r}|^2\xi^2}{(1+\mathcal{B} \xi^2)^2}\right)\\
&\equiv \frac{2}{\pi \ell^2_{\text{WC}}}\exp \left(-2|\bm{r}|^2/\ell^2_{\text{WC}}\right)
\end{split}
\end{equation}
where 
\begin{equation}
\begin{split}
\ell_{\text{WC}} \equiv \frac{1+\mathcal{B}\xi^2}{\xi}
\end{split}
\end{equation}
For small $\ell_{\text{WC}}$, the overlap between Wannier function localized on different sites become unimportant, and the interaction energy  per particle for the WC can be written as
\begin{equation}
\begin{split}
\frac{E^{\text{int}}_{\text{WC}}}{N}=&\int \frac{1}{2}d\bm{r}_1 d\bm{r}_2\sum_{\bm{R}\neq 0} n_0(\bm{r}_1)V(\bm{r}_1-\bm{r}_2)n_0(\bm{r}_2-\bm{R})\\
=&\sum_{\bm{R}\neq 0}\frac{1}{2}\frac{1}{4\pi^2}\int (n_0(\bm{k}))^2 V(\bm{k}) d\bm{k} \exp \left(i\bm{k}\cdot \bm{R}\right)\\
=&\sum_{\bm{R}\neq 0}\frac{V_c}{4\sqrt{\pi}\ell_{\text{WC}}}\exp \left(-\bm{|R|}^2/2\ell_{\text{WC}}^2\right)I_0(|\bm{R}|^2/2\ell_{\text{WC}}^2)
\end{split}
\end{equation}
This energy is the potential energy for a charge wave packet localized at $\bm{R}=\bm{0}$ under the potential produced by all other wave packets localized at  $\bm{R}\neq\bm{0}$. Note that when $|\bm{R}|$ is large, \begin{equation}
    \exp \left(-\bm{|R|}^2/2\ell_{\text{WC}}^2\right)I_0(|\bm{R}|^2/2\ell_{\text{WC}}^2)\to \frac{\ell_{\text{WC}}}{|\bm{R}|},
\end{equation}
which is divergent under summation $\sum_{\bm{R}\neq \bm{0}}$. As we shall show below, this divergence is associated with the charging energy of the system. We will need to subtract off this contribution, as we did for the IAHC and the FAHC. With this in mind, we expand the modified Bessel function around $|\bm{R}|\to \infty$
\begin{equation}\label{eq:madlungsum}
I_0(|\bm{R}|^2/2\ell_{\text{WC}}^2)=\frac{\exp \left(|\bm{R}|^2/2\ell_{\text{WC}}^2\right)}{\sqrt{\pi}}(\frac{\ell_{\text{WC}}}{|\bm{R}|}+\frac{\ell^3_{\text{WC}}}{4|\bm{R}|^3}+\frac{9\ell^5_{\text{WC}}}{32|\bm{R}|^5}+\frac{75\ell^7_{\text{WC}}}{128|\bm{R}|^7}+...)
\end{equation}
Such that 
\begin{equation}\label{eq:EWC}
\frac{E^{\text{int}}_{\text{WC}}}{N}=\sum_{\bm{R}\neq 0}\frac{V_c}{4\pi}(\frac{1}{|\bm{R}|}+\frac{\ell^2_{\text{WC}}}{4|\bm{R}|^3}+\frac{9\ell^4_{\text{WC}}}{32|\bm{R}|^5}+\frac{75\ell^6_{\text{WC}}}{128|\bm{R}|^7}+...)
\end{equation}

The divergence is from $\sum_{\bm{R}\neq \bm{0}}\frac{1}{|\bm{R}|}$. To evaluate this, we define the following function
\begin{equation}
h(\bm{r})\equiv\sum_{\bm{R}}\frac{1}{|\bm{r}-\bm{R}|}-\frac{1}{|\bm{r}|}
\end{equation}
Formally, $\sum_{\bm{R}\neq \bm{0}}\frac{1}{|\bm{R}|}=h(\bm{0})$. By Poission summation, 
\begin{equation}
h(\bm{r})=\sum_{\bm{g}}\frac{2\pi}{|\bm{g}|}\frac{1}{A_{uc}}\exp \left(i\bm{g}\cdot \bm{r}\right)-\frac{1}{|\bm{r}|}
\end{equation}
where the sum is over reciprocal lattice vectors $\bm{g}$, and $A_{uc}$ is the unit cell area corresponding to lattice $\bm{R}$. The divergence of $h(\bm{0})$ is from the $\bm{g}= 0$ term, which we will subtract off from $h(\bm{0})$ later. To proceed, we notice,
\begin{equation}
\begin{split}
&h(\bm{r})+\frac{1}{|\bm{r}|}=\sum_{\bm{R}}\frac{1}{|\bm{r}-\bm{R}|}\\
&=\sum_{\bm{R}}\frac{2}{\sqrt{\pi}}\int_{0}^{\infty} du \exp \left(-u^2 |\bm{r}-\bm{R}|^2\right)\\
&=\sum_{\bm{g}}\exp \left(i\bm{g}\cdot \bm{r}\right)\frac{2\sqrt{\pi}}{A_{uc}}\int^{\infty}_{0} \frac{du}{u^2} \exp \left(-|\bm{g}|^2/(4u^2)\right)\\
&=\sum_{\bm{g}}\exp \left(i\bm{g}\cdot \bm{r}\right)\frac{2\sqrt{\pi}}{A_{uc}}\int^{\epsilon}_{0} \frac{du}{u^2} \exp \left(-|\bm{g}|^2/(4u^2)\right)+\sum_{\bm{R}}\frac{2}{\sqrt{\pi}}\int_{\epsilon}^{\infty} du \exp \left(-u^2 |\bm{r}-\bm{R}|^2\right)\\
&=\sum_{\bm{g}}\exp \left(i\bm{g}\cdot \bm{r}\right)\frac{2\pi}{A_{uc}}\frac{\text{erfc}(\frac{|\bm{g}|}{2\epsilon})}{|\bm{g}|}+\sum_{\bm{R}}\frac{\text{erfc}(\epsilon |\bm{r}-\bm{R}|)}{|\bm{r}-\bm{R}|},
\end{split}
\end{equation}
where we used Poission summation to go from the second to the third line, and split the integral into two parts to go from the third to the fourth line. $\epsilon$ is an arbitrary parameter that allows for efficient numerical computation.  
The choice of $\epsilon$ does not affect the final result. $\text{erfc}(x)$ is the complementary error function. This is known in the literature as the Ewald summation (See, for example, SM of Ref.\cite{PhysRevLett.129.227202} for a detailed discussion). Subtracting off the $\bm{|g|}=\bm{0}$ term leads to
\begin{equation}
h(\bm{r})-\frac{2\pi}{|\bm{g}|A_{uc}}|_{\bm{g}=\bm{0}}=\sum_{\bm{g}\neq 0}\exp \left(i\bm{g}\cdot \bm{r}\right)\frac{2\pi}{A_{uc}}\frac{\text{erfc}(\frac{|\bm{g}|}{2\epsilon})}{|\bm{g}|}-\frac{2\sqrt{\pi}}{\epsilon A_{uc}}+\sum_{\bm{R}\neq 0} \frac{\text{erfc}(\epsilon |\bm{r}-\bm{R}|)}{|\bm{r}-\bm{R}|}-\frac{2\epsilon}{\sqrt{\pi}}
\end{equation}
This equation will quickly converge for appropriately chosen $\epsilon$, usually chosen to be $(\text{lattice constant}^{-1})$. The final expression for properly regularized $h(\bm{0})$ is
\begin{equation}
h(\bm{0})-\frac{2\pi}{|\bm{g}|A_{uc}}|_{\bm{g}=\bm{0}}=\frac{-4.21342}{a_0}
\label{eq:SubtractedDiv}\end{equation}
where $a_0$ is the lattice constant. This constant is associated with the Madelung energy of the Wigner Crystal. Other terms in the summation in Eq.\ref{eq:EWC} are convergent. In fact, there is a closed-form formula for these terms \cite{zucker1974note}
\begin{equation}
\sum^{m,n\in \mathcal{Z}}_{(m,n)\neq (0,0)}(m^2+n^2+mn)^{-s}=6\zeta(s)g(s)\equiv A(s) \qquad g(s)\equiv \sum_{n=0}^{\infty} (3n+1)^{-s}-(3n+2)^{-s}
\end{equation}
where $\zeta(s)$ are Riemann zeta function. Strictly speaking, this formula is only defined for $s>1/2$. But taking $s=1/2$ in the above formula will produce the correctly regularized answer, such that $A(\frac{1}{2})/a_0$ is equal to the right-hand side of Eq.~\ref{eq:SubtractedDiv}. Using this, we find the following expression for the regularized interaction energy  $E^{\text{int}}_{\text{WC}}$
\begin{equation}
\frac{E^{\text{int}}_{\text{WC}}}{N}=\frac{V_{\text{C}}}{4\pi}(\frac{A(\frac{1}{2})}{a_0}+\frac{A(\frac{3}{2})\ell^2_{\text{WC}}}{4a_0^3}+\frac{9A(\frac{5}{2})\ell^4_{\text{WC}}}{32a_0^5}+\frac{75A(\frac{7}{2})\ell^6_{\text{WC}}}{128a_0^7}+...).
\end{equation}
The divergent term we have subtracted off from $\frac{E^{\text{int}}_{\text{WC}}}{N}$ is
\begin{equation}
\frac{V_c}{4\pi}\frac{2\pi}{|\bm{q}|A_{uc}}|_{\bm{q}=\bm{0}}= \frac{\rho_0}{2} V(\bm{q}) |_{\bm{q}=\bm{0}}=\frac{\rho_0}{2} \int d\bm{r}V(\bm{r}).
\end{equation}
This is the same charging energy that we also subtracted off the energy of the IAHC, and FAHC, as discussed above. 

The kinetic energy of the Wigner Crystal is easy to calculate using the Wannier representation.  For large $\frac{1}{\Omega \ell^2_{\text{WC}}}$, the overlap between Wannier orbitals centered around different sites $\bm{R}$ is unimportant. In this limit, the kinetic energy per particle is
\begin{equation}
\frac{E^{\text{kin}}_{\text{WC}}}{N}=-\sum_{n}\frac{1}{2m_0}\int d\bm{r} \phi_n^*(\bm{r})\nabla^2  \phi_n(\bm{r})=\frac{\xi^2}{m_0}.
\end{equation}

\subsection{Integer Anomalous Hall Crystal}\label{sup:WannierIAHC}
We now return to the interaction and kinetic energy per particle of the IACH wavefunction using the Coulomb interaction in Eq.~\ref{eq:CInteraction}. 
Following the argument given in Ref.~\cite{tan2024parent}, total energy of the IAHC is equal to the Fock and kinetic energies of an appropriately chosen $C = 0$ wavefunction with a modified interaction. Here, the $C = 0$ wavefunction is the Slater determinant of the single particle states
\begin{equation}
\tilde{\psi}^{(1)}_{\bm{k},n}(\bm{r})=\sum_{\bm{g}}\mathcal{N}_{\bm{k}}\exp \left(-\frac{|\bm{k}+\bm{g}|^2}{4\xi^2}\right) \tilde{w}_{\bm{k}+\bm{g},n}(\bm{r}),
\end{equation}
where $\tilde{w}$ is defined as in Eq.~\ref{eq:ICBStates}, but with parent Berry curvature $\tilde{\mathcal{B}}\equiv\mathcal{B}-\frac{2\pi}{\Omega}=\delta$,
i.e. $\tilde{w}_{\bm{k},n}(\bm{r}) = \exp \left (i \bm{k}\cdot \bm{r} \right ) \braket{n|s^{\tilde{\mathcal{B}}}_{\bm{k}}}$.
The effective interaction is \footnote{This might seem problematic at $\delta<0$, as there $\braket{s^{\mathcal{B}'}_{\bm{k}}|s^{\mathcal{B}'}_{\bm{k}+\bm{q}}}$ diverges at large $|\bm{q}|$. However, this divergence is cancelled by the factor $\exp \left(-\frac{\pi|\bm{q}|^2}{\Omega}\right)$ in the effective interaction.}

\begin{equation}
\begin{split}
&\tilde{V}(\bm{q})=\frac{V_{\text{C}}}{ |\bm{q}|}\exp (-\frac{\pi |\bm{q}|^2}{\Omega})\\
&\tilde{V}(\bm{r})=\frac{1}{4\pi^2}\int \tilde{V}(\bm{q})\exp (i\bm{q}\cdot \bm{r}) d\bm{q}=\frac{V_c\sqrt{\Omega}}{4\pi}\exp (-\frac{\Omega|\bm{r}|^2}{8\pi})I_0 (\frac{\Omega|\bm{r}|^2}{8\pi})
\end{split}
\end{equation}
where $I_0$ is the modified Bessel function of the first kind. 

Since the $\tilde{\psi}^{(1)}_{\bm{k},n}$ wavefunction has zero Chern number, it can be Wannierized. The corresponding Wannier wavefunction $\tilde{\varphi}_{\bm{R},n}$ are defined via
\begin{equation}
\tilde{\psi}^{(1)}_{\bm{k},n}(\bm{r})=\sum_{\bm{R}}\frac{1}{\sqrt{N}} \exp \left(i\bm{k}\cdot \bm{R}\right)  \tilde{\varphi}_{\bm{R},n}(\bm{r})
\end{equation}
where $N$ is the number of unit cells (which is also equal to the number of electrons in this case), and $\bm{R}$ is summed over the real space lattice vectors. Using translation symmetry, we can define $\tilde{\varphi}_{\bm{R},n}(\bm{r})\equiv \tilde{\varphi}_n(\bm{r}-\bm{R})$, where, explicitly,
\begin{equation}
\begin{split}
\tilde{\varphi}_n(\bm{r})&=\frac{\sqrt{N}}{\Omega} \int  d \bm{q} \mathcal{N}_{\lceil \bm{q} \rceil} \exp \left(-\frac{|\bm{q}|^2}{4\xi^2}\right)  \tilde{w}_{\bm{q},n}.
\end{split}
\end{equation}
Similar to before, the charge density $\tilde{n}(\bm{r})$ associated with the $\tilde{\varphi}_n(\bm{r})$ is,
\begin{equation}
\begin{split}
\tilde{n}(\bm{r})&=\sum_{n}|\tilde{\varphi}_n(\bm{r})|^2\\
&=\frac{N}{\Omega^2}\int d\bm{q}_1 d\bm{q}_2   \mathcal{N}_{\lceil \bm{q}_1 \rceil}\mathcal{N}_{\lceil \bm{q}_2 \rceil} \exp \left(-(\bm{q}_1^2+\bm{q}_2^2)/4\xi^2\right)  \exp \left(i(\bm{q}_1-\bm{q}_2)\cdot\bm{r}\right)\exp \left(-\frac{\delta}{4}(|\bm{q}_1 - \bm{q}_2|^2 + 2i \bm{q}_1 \times \bm{q}_2 )\right)\\
&=\frac{2}{\pi}\frac{\xi^2}{(1+\delta\xi^2)^2}\exp \left(-\frac{2|\bm{r}|^2\xi^2}{(1+\delta \xi^2)^2}\right)\\
&\equiv \frac{2\xi_1^2}{\pi}\exp \left(-2|\bm{r}|^2\xi_1^2\right)
\end{split}
\end{equation}
where we have used Eq.~\ref{eq:NormalizationDef}, in the limit of small $\epsilon$, and that the system area is $A = \frac{4\pi^2 N}{\Omega}$. From this, we see that the Wannier function describes exponentially localized charge wave packets, with localization length $\xi^{-1}_1$. For large $\frac{\xi^2_1}{\Omega}$, the overlap between Wannier orbitals centered around different sites $\bm{R}$ is unimportant. In this limit, the Fock energy is equal to the minus the self-interaction of each wave packet $\phi_{\bm{R},n}(\bm{r})$\cite{tan2024parent}. Thus, the Fock energy per particle is
\begin{equation}
\begin{split}
\frac{E^{\text{int}}_{\text{IAHC}}}{N}&=-\int d\bm{r}_1 d\bm{r}_2\frac{1}{2} \tilde{n}(\bm{r}_1) \tilde{V}(\bm{r}_1-\bm{r}_2) \tilde{n}(\bm{r}_2)\\
&=-\frac{1}{2}\int \frac{1}{4\pi^2}(\tilde{n}(\bm{k}))^2 \tilde{V}(\bm{k}) d\bm{k}\\
&=-\frac{V_{\text{C}}}{4 \sqrt{\pi}  \sqrt{\frac{1}{\xi_1 ^2}+\frac{4\pi}{\Omega }}}.
\end{split}
\end{equation}
Similar to before, in the limit of large $\frac{\xi^2_1}{\Omega}$, the kinetic energy per particle is simply
\begin{equation}
\frac{E^{\text{kin}}_{\text{IAHC}}}{N}=-\sum_{n}\frac{1}{2m}\int d\bm{r} \tilde{\varphi}_n^{ *}(\bm{r})\nabla^2  \tilde{\varphi}_n(\bm{r})=\frac{\xi^2}{m_0}.
\end{equation}
Note that the kinetic and interaction energy obtained in this approach is equal to that obtained in the previous section.

\section{Integer Anomalous Hall Crystal Wavefunction From Functional Minimization}
The Wannier approach used in Sec.~\ref{sup:WannierIAHC}, also allows us to show that the IAHC wavefunction is a local minimum of the energy in certain limits. 
To do this, let us consider an IAHC made up of the following single-particle trial wavefunctions,
\begin{equation}
    \psi_{\bm{k},n}(\bm{r})=\sum_{\bm{g}} f(\bm{k}+\bm{g})\exp \left(-i\pi  (\frac{\bm{k}\times \bm{g}}{\Omega}+\omega_{\bm{g}})\right) w_{\bm{k}+\bm{g},n}(\bm{r})
\end{equation}
where $f(\bm{k})$ is a smooth function. These functions define an IAHC with Chern number $C = 1$. We will show that the energy of this state is (locally) minimized when $f(\bm{k})\propto \exp \left(-\frac{|\bm{k}|^2}{4\xi^2}\right)$. As before, we first map the $C = 1$ wavefunctions onto Chern number $C = 0$ wavefunctions
\begin{equation}
\tilde{\psi}_{\bm{k},n}(\bm{r})=\sum_{\bm{g}}f(\bm{k}+\bm{g})\tilde{w}_{\bm{k}+\bm{g},n}(\bm{r}).
\end{equation}
The Wannier representation of the $C = 0$ wavefunctions is
\begin{equation}
\tilde{\varphi}_n(\bm{r}-\bm{R})=\tilde{\varphi}_{n,\bm{R}}(\bm{r})=\frac{1}{\sqrt{N}}\sum_{\bm{k}}\exp \left(-i\bm{k}\cdot \bm{R}\right)\tilde{\psi}_{\bm{k},n}(\bm{r})= \frac{A}{4\pi^2\sqrt{N}}\int d\bm{k} f(\bm{k})\exp \left(-i\bm{k}\cdot \bm{R}\right)\tilde{w}_{\bm{k},n}(\bm{r}),
\end{equation}
and the Fourier transform of the Wannier functions $\tilde{\varphi}(\bm{k})$ is
\begin{equation}
\begin{split}\label{eq:WannierFTDef}
\tilde{\varphi}_n(\bm{r})=\int \frac{d\bm{k}}{4\pi^2} \tilde{\varphi}(\bm{k})\exp(i\bm{k}\cdot\bm{r})\braket{n|s^{\tilde{\mathcal{B}}}_{\bm{k}}}\\
\tilde{\varphi}(\bm{k}) = \sum_n \int d\bm{r}\braket{s^{\mathcal{\tilde{B}}}_{\bm{k}}|n} \tilde{\varphi}_n(\bm{r})\exp \left(-i\bm{k}\cdot \bm{r}\right) = \frac{A}{\sqrt{N}} f(\bm{k})
\end{split}
\end{equation}
Let us assume that the overlap between different Wannier functions, is small and can be ignored. Under this assumption, the energy of the trial IAHC wavefunction can be written as the sum of the kinetic and Fock energy evaluated in the Wannier basis, as we did in the previous section.
 \begin{equation}\label{eq:energyfunctional}
 \frac{E}{N}=\sum_{n}\int\frac{1}{2m_0} d\bm{r}(\nabla\tilde{\varphi}_n(\bm{r}))\cdot (\nabla\tilde{\varphi}_n^{*}(\bm{r}))-\frac{1}{2}\sum_{m,n}\int d\bm{r}d\bm{r'} \tilde{V}(\bm{r}-\bm{r}') |\tilde{\varphi}_n(\bm{r})|^2|\tilde{\varphi}_m(\bm{r}')|^2\equiv \frac{E^{\text{kin}}}{N} + \frac{E^{\text{int}}}{N}
 \end{equation}
Minimizing the energy functional with respect to small changes in $\tilde{\varphi}(\bm{k})$ yields the following,
\begin{equation}
\begin{split}
\frac{\delta E^{\text{kin}}}{N \delta \tilde{\varphi}^*(\bm{k})}
&=\frac{1}{2m_0}\int d\bm{r} d\bm{k}_1 d\bm{k}_2 (\bm{k}_1 \cdot\bm{k}_2) \tilde{\varphi} (\bm{k}_1) \delta(\bm{k}_2-\bm{k})\frac{1}{(4\pi^2)^2}\exp \left(i(\bm{k}_1-\bm{k}_2\right)\cdot \bm{r})\braket{s^{\mathcal{\tilde{B}}}_{\bm{k}_2}|s^{\mathcal{\tilde{B}}}_{\bm{k}_1}}\\
&=\frac{|\bm{k}|^2}{2m_0}\tilde{\varphi}(\bm{k})\frac{1}{4\pi^2},
\end{split}
\end{equation}
\begin{equation}
\begin{split}
\frac{\delta E^{\text{int}}}{N\delta \tilde{\varphi}^*(\bm{k})}&=-\sum_m \int d\bm{r}d\bm{r}' d\bm{k}_1 d\bm{k}_2|\tilde{\varphi}_m(\bm{r}')|^2 \tilde{V}(\bm{r}-\bm{r}') \frac{1}{(4\pi^2)^2} \tilde{\varphi}(\bm{k}_1)\delta(\bm{k}_2-\bm{k})\exp \left(i(\bm{k}_1-\bm{k}_2\right)\cdot \bm{r})\braket{s^{\mathcal{\tilde{B}}}_{\bm{k}_2}|s^{\mathcal{\tilde{B}}}_{\bm{k}_1}}\\
&=-\sum_m \int d\bm{r}d\bm{r}' d\bm{k}_1 |\tilde{\varphi}_m(\bm{r}')|^2 \tilde{V}(\bm{r}-\bm{r}') \frac{1}{(4\pi^2)^2} \tilde{\varphi}(\bm{k}_1)\exp \left(i(\bm{k}_1-\bm{k})\cdot \bm{r}\right)\braket{s^{\mathcal{\tilde{B}}}_{\bm{k}}|s^{\mathcal{\tilde{B}}}_{\bm{k}_1}}\\
\label{eq:IntEnergyDeriv}\end{split}\end{equation}
To proceed, we set, by fiat, $\sum_m |\tilde{\varphi}_m(\bm{r}')|^2 = \tilde{n}(\bm{r}') = \frac{2\xi_1^2}{\pi}\exp \left(-2|\bm{r}'|^2\xi_1^2\right)$. We will show later that this leads to a self-consistent solution to the minimization equations. Using this form of $\tilde{n}(\bm{r})$, 
\begin{equation}\label{eq:trapping}
\begin{split}
\int \sum_m d\bm{r'} |\tilde{\varphi}_m(\bm{r}')|^2 \tilde{V}(\bm{r}-\bm{r'})&=\int d\bm{r} \tilde{n}(\bm{r'})\tilde{V}(\bm{r}-\bm{r'})=\int \frac{1}{4\pi^2} d\bm{k}\tilde{n}(\bm{k}) \tilde{V}(\bm{k})\exp \left(i\bm{k}\cdot\bm{r}\right)\\ &=\frac{V_c \sqrt{\Omega_e}}{4\pi}\exp \left(-\frac{\Omega_e |\bm{r}|^2}{8\pi}\right) I_0(\frac{\Omega_e |\bm{r}|^2}{8\pi})\approx \frac{V_c\sqrt{\Omega_e}}{4\pi}-\beta |\bm{r}|^2,
\end{split}
\end{equation}
where we defined \begin{equation}
\frac{\pi}{\Omega_e}\equiv\frac{\pi}{\Omega}+\frac{1}{8\xi_1^2}, \quad\quad \beta\equiv \frac{V_C\sqrt{\Omega_e^3}}{32\pi^2},
\end{equation}
and taken $|\bm{r}|$ small compared to $\Omega_e^{-\frac{1}{2}}$, an assumption that will be confirmed self-consistently. From this, we can reduce Eq.~\ref{eq:IntEnergyDeriv} to
\begin{equation}
    \begin{split}
\frac{\delta E^{\text{int}}}{N\delta \tilde{\varphi}^*(\bm{k})}&=\int d\bm{r} d\bm{k}_1 \exp \left(i(\bm{k}_1-\bm{k})\cdot \bm{r}\right)\braket{s^{\mathcal{\tilde{B}}}_{\bm{k}}|s^{\mathcal{\tilde{B}}}_{\bm{k}_1}}\left [ \beta \frac{|\bm{r}|^2}{(4\pi^2)^2} - \frac{V_c\sqrt{\Omega_e}}{4\pi(4\pi^2)^2}  \right ]\tilde{\varphi}(\bm{k}_1)\\
&=\sum_n \beta\int d\bm{r}  \frac{|\bm{r}|^2}{4\pi^2} \tilde{\varphi}_n(\bm{r})\exp \left(-i\bm{k}\cdot \bm{r}\right)\braket{s^{\mathcal{\tilde{B}}}_{\bm{k}}|n}-
\sum_n \frac{V_c\sqrt{\Omega_e}}{4\pi}\int d\bm{r}  \frac{1}{4\pi^2} \tilde{\varphi}_n(\bm{r})\exp \left(-i\bm{k}\cdot \bm{r}\right)\braket{s^{\mathcal{\tilde{B}}}_{\bm{k}}|n}\\
&=-\sum_n \beta \braket{s^{\mathcal{\tilde{B}}}_{\bm{k}}|n} \frac{1}{4\pi^2} \nabla^2_{\bm{k}}[\tilde{\varphi}(\bm{k})\braket{n|s^{\mathcal{\tilde{B}}}_{\bm{k}}}]- \frac{V_c\sqrt{\Omega_e}}{4\pi}\frac{1}{4\pi^2}\tilde{\varphi}(\bm{k}).
\end{split}
\end{equation}
To summarize, the minimization equation for $\tilde{\varphi}(\bm{k})$ is
\begin{equation}
\frac{|\bm{k}|^2}{2m_0} \tilde{\varphi}(\bm{k})-\sum_n \beta \braket{s^{\mathcal{\tilde{B}}}_{\bm{k}}|n} \nabla^2_{\bm{k}}[\tilde{\varphi}(\bm{k})\braket{n|s^{\mathcal{\tilde{B}}}_{\bm{k}}}]=\varepsilon \tilde{\varphi}(\bm{k})
\label{eq:MinEq}\end{equation}
subject to the additional self-consistency condition that $\tilde{n}(\bm{r}) = \frac{2\xi_1^2}{\pi}\exp \left(-2|\bm{r}|^2\xi_1^2\right)$. Here, $\varepsilon$ is the Lagrangian multiplier\footnote{We combine the constant term when doing the harmonic approximation in Eq.\ref{eq:trapping} with the multiplier} from requiring  $\int d\bm{k}|\tilde{\varphi}(\bm{k})|^2$ to be conserved. 
Note that the self-consistency condition also guarantees that the wavefunction we find is properly normalized.  
We now make use of the form factors as in Eq.\ref{eq:FormFactSM}, 
\begin{equation}
\frac{|\bm{k}|^2}{2m_0} \tilde{\varphi}(\bm{k})- \beta  [(\nabla_{\bm{k}}-i\tilde{\bm{A}})^2 \tilde{\varphi}(\bm{k})-\tilde{\mathcal{B}} \tilde{\varphi}(\bm{k})]=\varepsilon \tilde{\varphi}(\bm{k}) 
\quad  \tilde{\bm{A}}=\frac{\tilde{\mathcal{B}}}{2}(-k_y,k_x)
\label{eq:FDHam}\end{equation}
This is the Schrödinger equation for two-dimensional harmonic oscillator in magnetic field, also known as the Fock-Darwin Hamiltonian. 
The lowest energy eigenstate of the Fock-Darwin Hamiltonian is well known to be Gaussian \cite{darwin1931diamagnetism,fock1928bemerkung}. 
Recall from earlier that $\tilde{n}(\bm{r}')= \frac{2\xi_1^2}{\pi}\exp \left(-2|\bm{r}'|^2\xi_1^2\right)$. Because of this, we must have $\tilde{\varphi}(\bm{k})\propto \exp \left(-\frac{|\bm{k}|^2}{4\xi^2}\right)$ (see Eq. \ref{eq:WannierFTDef}) in order for the solution of the minimization equation to be self-consistent. Using this form of $\tilde{\varphi}(\bm{k})$ with Eq.~\ref{eq:FDHam},

\begin{equation}
\frac{|\bm{k}|^2}{2m_0}\exp \left(-\frac{|\bm{k}|^2}{4\xi^2}\right)+\frac{\beta}{4\xi^4}\exp \left(-\frac{|\bm{k}|^2}{4\xi^2}\right) (1+\tilde{\mathcal{B}}\xi^2)(4\xi^2+|\bm{k}|^2(-1+\tilde{\mathcal{B}}\xi^2))=\epsilon \exp \left(-\frac{|\bm{k}|^2}{4\xi^2}\right) 
\end{equation}
A solution to this equation will be a local minimum of the energy, provided that the small $|\bm{r}|$ approximation in Eq~\ref{eq:trapping} is valid.

In the limit $m_0 \to \infty$ (equivalently, $V_{\text{C}}\to \infty$ or $\beta \to \infty$), the solution is
\begin{equation}
\xi^2 =\frac{1}{|\tilde{\mathcal{B}}|}=\frac{1}{|\delta|}
\end{equation}
This is exactly the same $\xi$ we would obtain from directly minimizing the IAHC energy in the same limit. 
This serves as a good consistency check.
For finite $m_0$, the self-consistency equation is
\begin{equation}
\begin{split}
&\frac{1}{2m_0}+\frac{\beta}{4\xi^4}(1+\tilde{\mathcal{B}}\xi^2)(-1+\tilde{\mathcal{B}}\xi^2)=0\\
\to &\frac{1}{2m_0}+\frac{V_{\text{C}}}{4\sqrt{2\pi}\xi^4}(1+\delta \xi^2)(-1+\delta \xi^2) \sqrt{\frac{\xi^6 \Omega^3}{(8\pi\xi^2+(1+\delta \xi^2)^2\Omega)^3}}=0
\end{split}
\end{equation}
In the limit of large $\xi_1$, $\sqrt{\frac{\xi^6 \Omega^3}{(8\pi\xi^2+(1+\delta \xi^2)^2\Omega)^3}}\to \frac{1}{16\sqrt{2}}(\Omega/\pi)^{3/2}$, then the above equation further reduce to

\begin{equation}\label{eq:largexi1self}
\frac{1}{2m_0}+\frac{V_{\text{C}}}{128\pi^2\xi^4}(1+\delta \xi^2)(\delta \xi^2-1) \Omega^{3/2}=0
\end{equation}
On the other hand, if we directly minimize $E^{\text{kin}}_{\text{IAHC}}+E^{\text{int}}_{\text{IAHC}}=\frac{\xi^2}{m_0}-\frac{V_{\text{C}}}{4\sqrt{\pi}}\frac{1}{\sqrt{1/\xi_1^2+\frac{4\pi}{\Omega}}}$ with respect to $\xi$, the corresponding equation is
\begin{equation}
\frac{1}{2m_0}+\frac{V_{\text{C}}}{16\sqrt{\pi}\xi^4}\frac{(1+\delta \xi^2)(\delta \xi^2-1)}{(\frac{(1+\delta\xi^2)^2}{\xi^2}+\frac{4\pi}{\Omega})^{3/2}}=0
\end{equation}
This reduces to Eq. \ref{eq:largexi1self} in the limit of large $\xi_1$. Thus, the two approaches will agree at large $\xi_1$, where the Wannier functions are well localized. 
In this limit, our approximation of small $|\bm{r}|^2$ in Eq~\ref{eq:trapping} is also justified.
We note that $\xi_1$ is large near $\rho_0 = 1/2\pi \mathcal{B}$, and when interactions are strong (equivalently, when the effective mass is large). 
Finally, we remark that this proof can be easily generalized to the WC wavefunction. The only difference is in the energy functional Eq.\ref{eq:energyfunctional}, the interaction energy includes both Hartree and Fock contributions. Regardless, these interaction effects can still be approximated by a harmonic potential, as we did in Eq.\ref{eq:trapping}.

\section{Doping the IAHC}
Here, we address the outcome of adding a single hole to an IAHC. There are two main possibility to consider. (1) the period of the IAHC stays fixed, and the doped carrier is free (2) the IAHC distorts and reduces it period to accommodate the additional hole. Here, we will show that (2) occurs near $\rho_0 = 1/(2\pi \mathcal{B})$. To simplify analytic calculations, we consider the IAHC in the $m_0\rightarrow \infty$, limit. 

First, we consider the energy cost of adding a single hole. The total interaction energy of the IAHC is
\begin{equation}
E^{\text{int}}_{\text{IAHC}}=\frac{1}{2}\sum_{\bm{k}_1}(\sum_{\bm{k}_2\neq \bm{k}_1} V_{\bm{k}_1\bm{k}_2\bm{k}_1\bm{k}_2}-V_{\bm{k}_1\bm{k}_2\bm{k}_2\bm{k}_1})\\
\end{equation}
where $V_{\bm{k}_1\bm{k}_2\bm{k}_3\bm{k}_4}$ are the matrix elements of the interaction Hamiltonian. In terms of the IAHC ansatz wavefunctions  $\psi^{(1)}_{\bm{k},n}$,
\begin{equation}
V_{\bm{k}_1\bm{k}_2\bm{k}_3\bm{k}_4}=\int d\bm{r}_1 d\bm{r}_2 V(\bm{r}_1-\bm{r}_2)\sum_{m,n}\psi^{(1)*}_{\bm{k}_1,m} (\bm{r}_1)\psi^{(1)}_{\bm{k}_3,m}(\bm{r}_1) \psi^{(1)*}_{\bm{k}_2,n}(\bm{r}_2) \psi^{(1)}_{\bm{k}_4,n}(\bm{r}_2). 
\end{equation}
The energy associated with doping a hole with momentum $\bm{q}$ is
\begin{equation}
E_{\text{dope}}=-(\sum_{\bm{k}_2\neq \bm{q}} V_{\bm{k}_1\bm{q}\bm{k}_1\bm{q}}-V_{\bm{k}_1\bm{q}\bm{q}\bm{k}_1}).
\end{equation}
Notice that the above equation is essentially Koopman's theorem \cite{szabo2012modern}. At strong interaction, this quantity should not depend on which orbital we choose to dope. This can be verified using the form factors of the ansatz Bloch states we derived earlier. The energy cost of doping one hole to the IAHC is therefore
\begin{equation}
E_{\text{dope}}= -2\frac{E^{\text{int}}_{\text{IAHC}}}{N}=\frac{V_{\text{C}}}{2\sqrt{\pi}}\frac{1}{\sqrt{2\mathcal{B}+2|\delta|}}.
\end{equation}
We now consider the energy cost associated with distorting the crystal lattice. To do this, we use the explicit form of the $m_0\rightarrow \infty$ IAHC energy in Eq.~\ref{eq:IAHCIntEnergy}. The density dependence is encoded in the parameter $\delta =  -\frac{A}{2\pi N} + \mathcal{B}$, where we have used that $\Omega = 4\pi^2 \rho_0 = \frac{4\pi^2 N }{A}$ for the IAHC. The energy cost with going from $N$ to $N-1$ particles in the IAHC is then
\begin{equation}
E_{\text{dist}}=-\frac{d E^{\text{int}}_{\text{IAHC}}}{dN}=
\begin{cases}
\frac{3 V_{\text{C}}}{8 \sqrt{2 \pi } \sqrt{\mathcal{B}-\delta }} \quad (\delta\leq 0)\\
\frac{V_{\text{C}} (\mathcal{B}+3 \delta )}{8 \sqrt{2 \pi } (\mathcal{B}+\delta )^{3/2}} \quad (\delta >0).
\end{cases}
\end{equation}
In either cases $E_{\text{dist}}<E_{\text{dope}}$. It is therefore energetically favorable for the IAHC to distort rather than letting the doped hole act as a free carrier. 

We explicitly note that in this analysis, we did not consider the other states that compete with IAHC, i.e., the Wigner crystal and the FAHC. As such, the above analysis is only applicable when the other states have significantly higher energy than the IAHC. As we have shown numerically, this is the case near density $\rho_0 = 1/2\pi \mathcal{B}$.

\section{Liquid Phase}
We estimate the energy of an uncorrelated Fermi Liquid (FL) state in the ICB. 
The FL state is defined as the Slater determinant state obtained by filling all states $\ket{\bm{k}} = \exp \left(i\bm{k}\cdot\bm{r}\right)\ket{s^{\mathcal{B}}_{\bm{k}}}$ with $\frac{|\bm{k}|^2}{2m_0}<E_F\equiv \frac{k^2_f}{2m_0}$, where $E_F$ is the Fermi energy, and $k_f = \sqrt{\frac{4\pi N}{A}}$ is the Fermi momentum.

The kinetic energy per particle for the FL is
\begin{equation}
\frac{E^{\text{kin}}_{\text{FL}}}{N}=\frac{1}{N}\sum_{\bm{k}}\frac{|\bm{k}|^2}{2m_0}=\frac{A}{4\pi^2 N}\int^{k_f}_{0}\frac{k^2}{2m_0}2\pi k dk=\frac{Ak_f^4}{16m_0 N\pi}=\frac{k_f^4}{16m_0 \rho_0 \pi}
\end{equation}
The interaction energy per particle for the FL is
\begin{equation}
\frac{E^{\text{int}}_{\text{FL}}}{N}=\frac{1}{2AN}\sum_{\bm{q}}\sum_{\substack{|\bm{k}_1|<k_f\\
|\bm{k}_2|<k_f}}\frac{V_{C}}{|\bm{q}|}\braket{s^{\mathcal{B}}_{\bm{k}_1-\bm{q}}|s^{\mathcal{B}}_{\bm{k}_1}}\braket{s^{\mathcal{B}}_{\bm{k}_2+\bm{q}}|s^{\mathcal{B}}_{\bm{k}_2}}[\delta_{\bm{q},\bm{0}}-\delta_{\bm{q},\bm{k}_1-\bm{k}_2}]
\end{equation}

\begin{figure}[t!]
    \centering
    \includegraphics[width=0.5\linewidth]{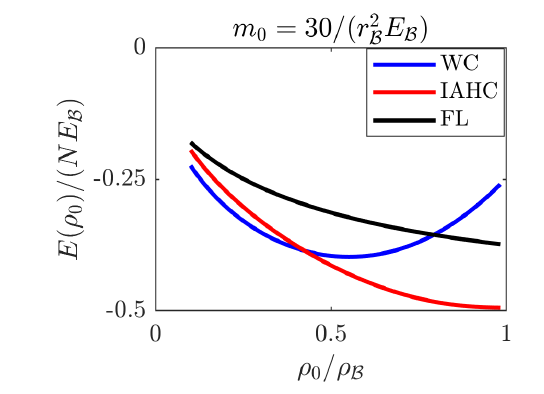}
    \caption{Energy per particle $E(\rho_0)/N\equiv (E_{\text{FL}}^{\text{int}}+E_{\text{FL}}^{\text{kin}})/N$ of Fermi Liquid compared with Wigner Crystal and Integer Anomalous Hall Crystal as a function of density $\rho_0/\rho_{\mathcal{B}}$ with $\rho_{\mathcal{B}}\equiv \frac{1}{2\pi r^2_{\mathcal{B}}}$}
    \label{SPfig_FL.pdf}
\end{figure}

The first term is the divergent charging energy, so we subtract it from the definition of  $E^{\text{int}}_{\text{FL}}$. After this subtraction, the modified $E^{\text{int}}_{\text{FL}}$ is,
\begin{equation}
\begin{split}
\frac{E^{\text{int}}_{\text{FL}}}{N}&=-\frac{1}{2AN}\sum_{\bm{q}}\sum_{\substack{|\bm{k}_1|<k_f\\
|\bm{k}_2|<k_f}}\frac{V_{C}}{|\bm{q}|}\braket{s^{\mathcal{B}}_{\bm{k}_1-\bm{q}}|s^{\mathcal{B}}_{\bm{k}_1}}\braket{s^{\mathcal{B}}_{\bm{k}_2+\bm{q}}|s^{\mathcal{B}}_{\bm{k}_2}}\delta_{\bm{q},\bm{k}_1-\bm{k}_2}\\
&=
-\frac{1}{2AN}\sum_{\substack{|\bm{k}_1|<k_f\\
|\bm{k}_2|<k_f}}\frac{V_{C}}{|\bm{k}_2-\bm{k}_1|}|\braket{s^{\mathcal{B}}_{\bm{k}_2}|s^{\mathcal{B}}_{\bm{k}_1}}|^2\\
&=-\frac{1}{2\rho_0 (4\pi^2)^2}\int^{|\bm{k}_1|<k_f}\int^{|\bm{k}_2|<k_f} d\bm{k}_1d\bm{k}_2 \frac{V_{C}}{|\bm{k}_2-\bm{k}_1|}\exp \left(-\frac{\mathcal{B}|\bm{k}_2-\bm{k}_1|^2}{2}\right)
\end{split}
\end{equation}
In Fig.\ref{SPfig_FL.pdf}, we show the energy of WC, IAHC and FL as a function of density, for $m_0r^2_{\mathcal{B}}E_{\mathcal{B}}=30$. We find that the FL is never the lowest energy state for all densities considered here. 
We note that the competition between the FL and the competing crystals will likely be much closer in reality due to correlation effects (as in usual 2D electron systems~\cite{PhysRevLett.102.126402}).

\section{Error Analysis}\label{sup:ErrAnalysis}
We made several approximations to find the energy of the FAHC. First, we ignored the $\bm{k}$-dependence of the normalization constant $\mathcal{N}_{\bm{k}}$. Second, we ignored all the terms with $\bm{R}\neq \bm{0}$ when deriving $F(\bm{k}+\bm{q},\bm{k})=\exp \left(-\frac{|\bm{q}|^2}{8\xi_1^2}\right)F_{LLL}(\bm{k}+\bm{q},\bm{k})$. Third, we made the same approximation when deriving the kinetic energy per particle is $\frac{\xi^2}{m_0}$. Under these assumptions, the energy of FAHC with Coulomb interaction $\frac{V_{\text{C}}}{ |\bm{q}|}$ is the same as the energy of FQH wavefunction with effective interaction $\frac{V_{\text{C}}}{|\bm{q}|}\exp \left(-\frac{|\bm{q}|^2}{4\xi_1^2}\right)$ plus a kinetic energy contribution, $\frac{\xi^2}{m_0}$. 

\begin{figure}[t!]
    \centering
    \includegraphics[width=0.98\linewidth]{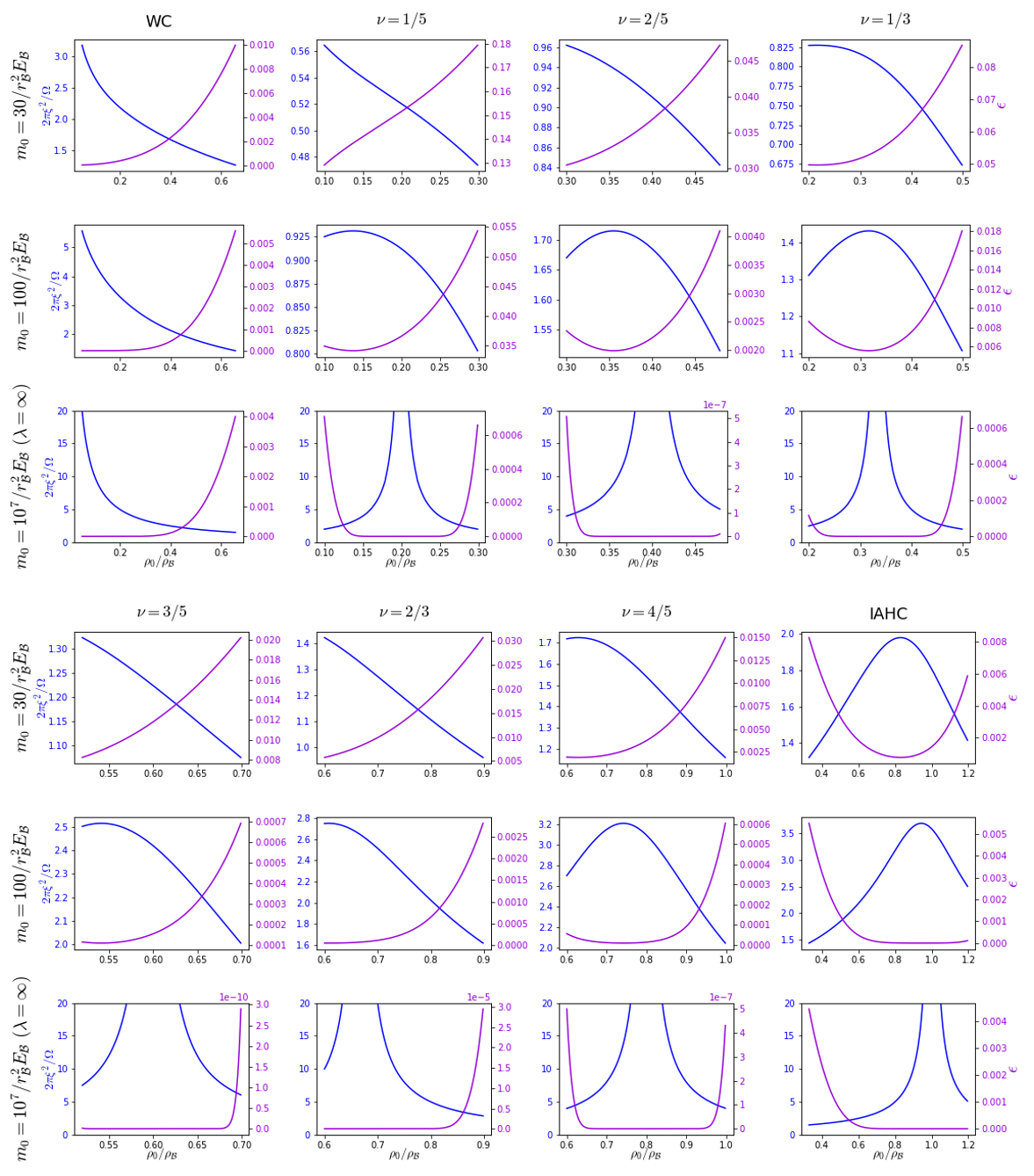}
    \caption{The varitionally optimized values of $2\pi \xi^2/\Omega$, and $\epsilon = \exp(-4\pi^2\xi^2/(\sqrt{3}\Omega))$ as a function of density, $\rho_0/\rho_{\mathcal{B}}$, for the different wavefunctions considered in the main text.}
    \label{fig:xiandep}
\end{figure}

We argued that the error associated with these approximations is unimportant when $\epsilon$ (Eq.~\ref{eq:epsilonDef}) is small (equiv. when $2\pi\xi^2/\Omega$ is significantly large). In Fig.~\ref{fig:xiandep} we show the variationally optimized values of $\epsilon$ and $2\pi\xi^2/\Omega$ for each state considered in the main text. We find that $\epsilon$ is very small for the FAHC states, when $m_0 = 10^7/r_\mathcal{B}^2 E_\mathcal{B}$. This is also the regime where the FAHC states are energetically favorable over a large range of densities. This indicates that the errors from our approximations are unimportant in the strongly correlated FAHC regime. For lower values of $m_0$, $\epsilon$ is larger. The errors induced by our approximation may therefore affect the competition between the FAHC, WC, and IAHC states at lower values of $m_0$. However, this will only change the details of the phase boundaries between different crystal states. The overall shape of the phase diagram we have constructed should be unchanged. 

To more fully characterize the error, we will also relate the values of $\epsilon$ (equiv. $2\pi\xi^2/\Omega$) to uncertainties in energy. We do this by comparing the following two energies: $E_1$, the energy of the FAHC calculated using the aforementioned assumptions for finite system size; $E_2$, the energy found using numeric diagonalized, projected onto the space of IAHC wavefunctions, $\ket{\bm{k}^{(1)}}$. Both energies are calculated using the same system size, number of particles, and the same fixed value of $2\pi\xi^2/\Omega$. In addition to the errors accumulated in the approximations above, this comparison also reveals the differences in energy between the FQH wavefunctions we used (e.g. the Laughlin state on the torus) and the true ground state with bare Coulomb interactions projected into the IAHC band.

For a fixed value of $\xi$, the value of $E_1$ is, explicitly,
\begin{equation}
\frac{E_1}{N}\equiv \frac{\braket{\Phi^{\text{FH}}|\hat{V}_1|\Phi^{\text{FH}}}}{N}+\frac{\xi^2}{m_0},
\end{equation}
where $\ket{\Phi^{\text{FH}}}$ is obtained from exact diagonalization within the LLL with the appropriate pseudo potential on a torus (with $l_B^2 = \frac{2\pi}{\Omega}$), and $\hat{V}_1$ is the pair interaction $V_1(\bm{q})=\frac{V_{\text{C}}}{ |\bm{q}|}\exp \left(-\frac{|\bm{q}|^2}{4\xi_1^2}\right)$. Note that the use of $\ket{\Phi^{\text{FH}}}$ in the calculation is, implicitly, the same approximation used in Eq.~\ref{eq:exactWF}.

$E_2$ is the ground state energy (found using numerical diagonalization) of the following Hamiltonian, defined in the Hilbert space spanned by Bloch state $\ket{\bm{k}^{(1)}}$,
\begin{equation}
\hat{\mathcal{H}}=\sum_{\bm{k}_1,\bm{k}_2\in BZ,\bm{q}}\frac{1}{2A} \frac{V_c}{|\bm{q}|} F(\bm{k}_1+\bm{q},\bm{k}_1)F(\bm{k}_2-\bm{q},\bm{k}_2) c^{(1)\dagger}_{\lceil \bm{k}_2-\bm{q} \rceil} c^{(1)\dagger}_{\lceil \bm{k}_1+\bm{q} \rceil} c^{(1) }_{\bm{k}_1} c^{(1) }_{\bm{k}_2}-\sum_{\bm{k}\in BZ}\braket{\bm{k}^{(1)}|\frac{\nabla^2}{2m_0}|\bm{k}^{(1)}} c^{(1) \dagger}_{\bm{k}} c^{(1) }_{\bm{k}}.
\end{equation}
Here, we do use any approximation for $F(\bm{k}+\bm{q},\bm{k})$ or $\braket{\bm{k}^{(1)}|\frac{\nabla^2}{2m_0}|\bm{k}^{(1)}} $, nor do we make any use of the wavefunction approximation in Eq.~\ref{eq:exactWF}. Note that the values of $\xi$ and $\Omega$ are encoded in the form factors. 

\begin{figure}[t!]
    \centering
    \includegraphics[width=0.98\linewidth]{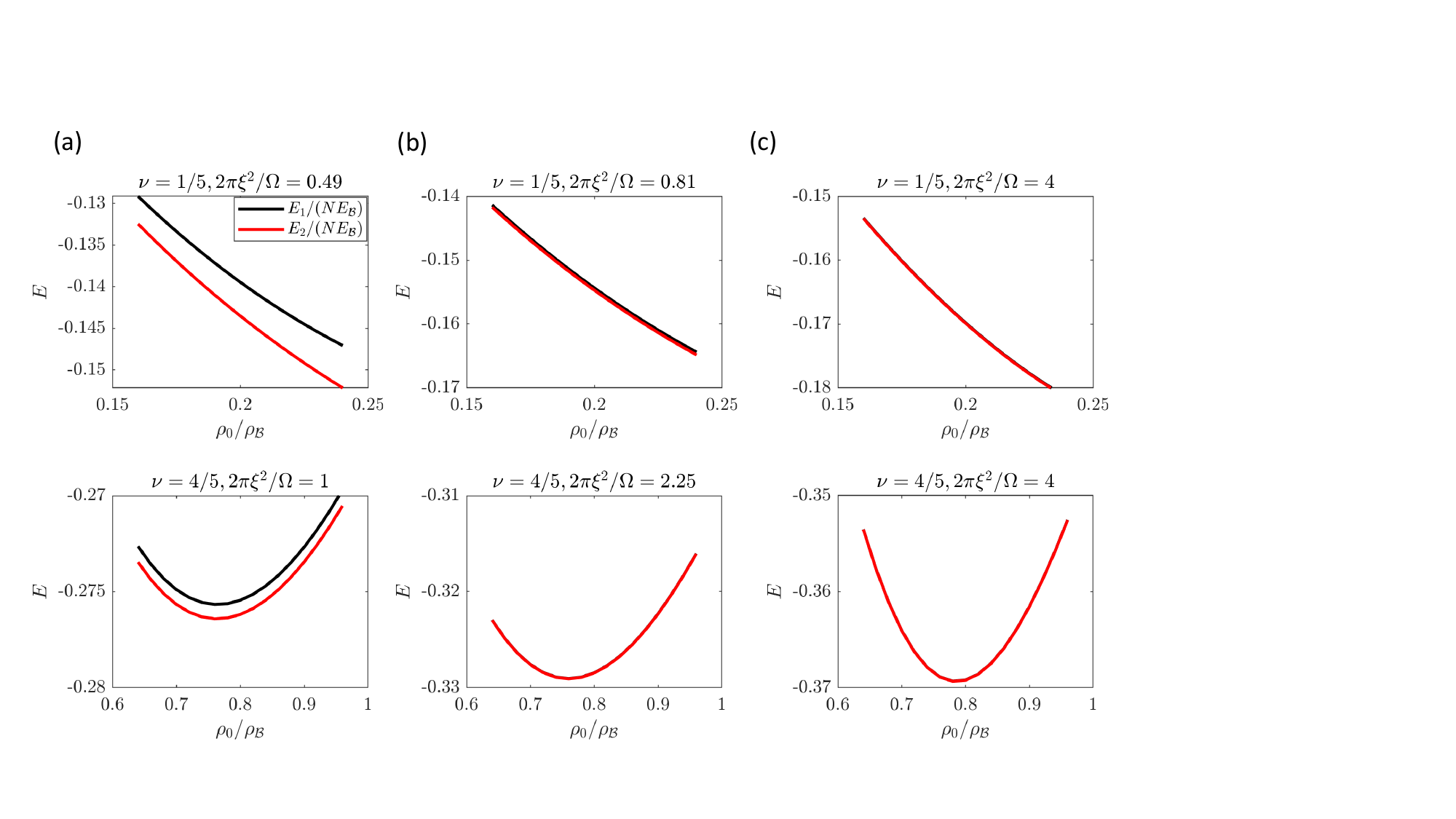}
    \caption{\textbf{(a)} $E_1/N$ and $E_2/N$ as a function of density $\rho_0/\rho_{\mathcal{B}}$. The maximum of difference between two lines in (a) is $5\times 10^{-3}$ (top),  $8\times 10^{-4}$ (bottom). $m_0=30/(r^2_{\mathcal{B} }E_{\mathcal{B}})$ \textbf{(b)} similar to (a), but for $m_0=100/(r^2_{\mathcal{B} }E_{\mathcal{B}})$. The maximum of the difference between the red line and black line is $4\times 10^{-4}$ (top),  $2 \times 10^{-5}$ (bottom). \textbf{(c)} similar to (a), but for $m_0=10^7/(r^2_{\mathcal{B} }E_{\mathcal{B}})$. The maximum of difference between two lines in (a) is $8\times 10^{-5}$ (top) and $2\times 10^{-5}$ (bottom).
    ED  is done on 5-by-5 25-site cluster. }
    \label{fig:SPfig_error}
\end{figure}

We show $E_1$ and $E_2$ for a selection of FAHC states in Fig.~\ref{fig:SPfig_error} at different fixed values of $2\pi \xi^2/\Omega$. We find that even for values of $2\pi \xi^2/\Omega$ that are lower than the lowest variationally optimized value in Fig~\ref{fig:xiandep}, the difference between $E_1$ and $E_2$ becomes vanishingly small at $m_0 = 10^7 / r_\mathcal{B}^2 E_\mathcal{B}$. This confirms the accuracy of our large $m_0$ results. The error in energy is larger at lower values of $m_0$. However, for the values of $m_0$ we consider, the error is not large enough to fundamentally change our results. 


\section{Extended Data}

\begin{figure}[t!]
    \centering
    \includegraphics[width=0.98\linewidth]{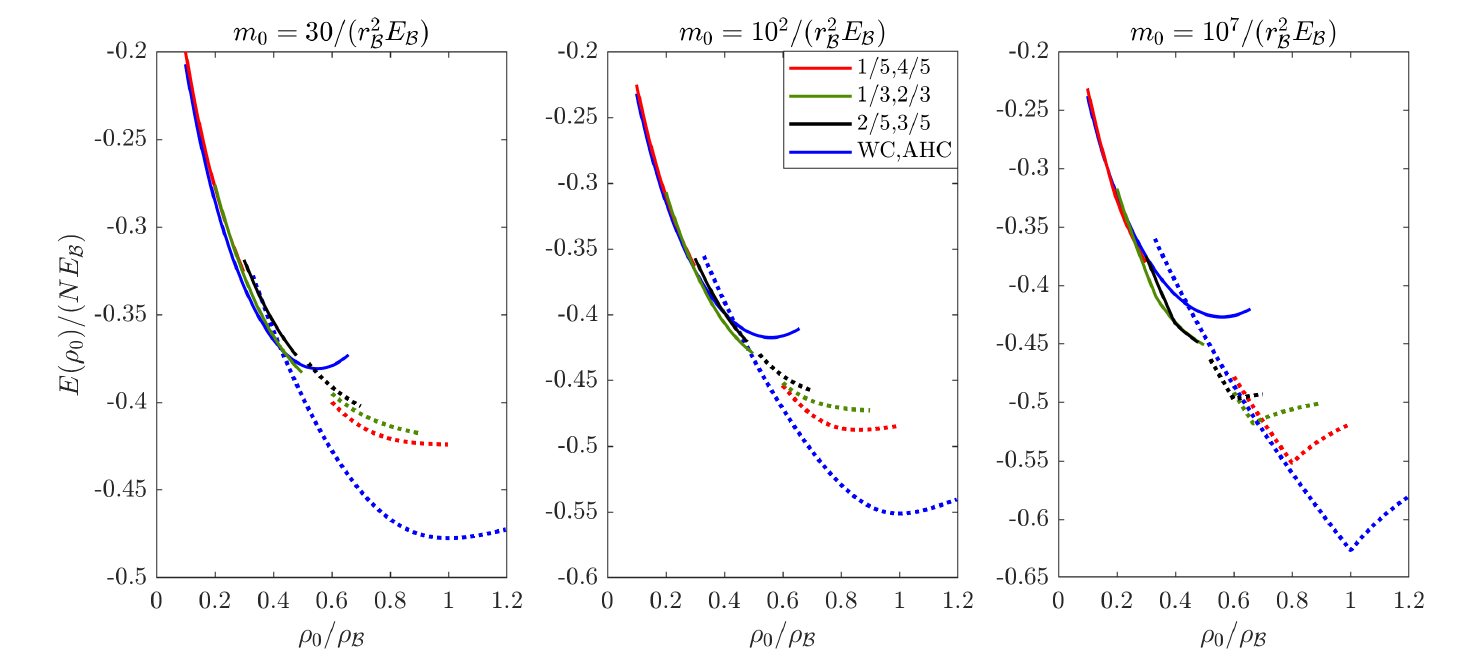}
    \caption{Energy per particle $E(\rho_0)/N E_{\mathcal{B}}$ of various candidate phases without subtracting $E_0(\rho_0)$. The left panel uses $m_0r^2_{\mathcal{B}} E_{\mathcal{B}}=30$, the middle panel uses $m_0r^2_{\mathcal{B}} E_{\mathcal{B}}=10^2$, the right panel uses $m_0r^2_{\mathcal{B}} E_{\mathcal{B}}=10^7 (\sim \infty)$. These are the same set of parameters as presented in the main text. 
    The solid lines represent the WC and $1/5, 1/3, 2/5$ FAHCs, while the dashed lines represent the IAHC and the $4/5, 2/3, 3/5$ FAHCs.
    }
    \label{SPfig_originaldata.pdf}
\end{figure}

The original data for the main text figure without subtracting $E_0(\rho_0)$ is shown in Fig.\ref{SPfig_originaldata.pdf}. $E_0(\rho_0)$ is obtained by fitting $\min\{E_{\text{WC}}(\rho_0), E_{\text{IC}}(\rho_0)\}$. We subtracted out $E_0(\rho_0)$ for better presentation. We use the following fitting formula for $E_0(\rho_0)$
\begin{equation}
E_0(\rho_0)/(NE_{\mathcal{B}})=A \rho+B\sqrt{\rho}+C\quad \rho\equiv \rho_0/\rho_{\mathcal{B}}
\label{eq:E0Fit}\end{equation}
The fitting results are give in Table~\ref{tb:fitresult2}.

\begin{center}
\begin{table}
\begin{tabular}{ |c|c|c|c| } 
\hline
\quad & $m_0 =30/ r^2_{\mathcal{B}}E_{\mathcal{B}}$ & $10^2/ r^2_{\mathcal{B}}E_{\mathcal{B}}$ & $10^7/r^2_{\mathcal{B}}E_{\mathcal{B}}\approx \infty$   \\
\hline
$A$ & $0.2770$ & $0.2288$ & $0.2019$   \\
\hline
$B$ & $-0.7754$ &$-0.7629$ & $-0.7496$  \\
\hline
$C$ & $-0.0238$ & $-0.0324$ & $-0.0889$   \\
\hline
\end{tabular}
\caption{Fitting coefficients for $E_0$ in Eq.~\ref{eq:E0Fit}.}\label{tb:fitresult2}
\end{table}
\end{center}


\end{document}